\documentclass[%
 amsmath,amssymb
]{revtex4-2}

\usepackage{graphicx}
\usepackage{bm}
\usepackage{stmaryrd}
\usepackage{float}
\usepackage{bbm}
\usepackage{amssymb}
\usepackage{algpseudocode}
\usepackage{algorithm} 
\usepackage[algo2e]{algorithm2e} 

\begin{document}

\preprint{}

\title{Isolating the hard core of phaseless inference: the Phase selection formulation}%

\author{Davide Straziota}
\author{Luca Saglietti}%

\affiliation{AI Lab, Computing Sciences department, Bocconi University.
}%

\date{\today}

\begin{abstract}
Real-valued Phase retrieval is an archetypal non-convex continuous inference problem, where a high-dimensional signal is to be reconstructed from an extensive dataset of signless linear measurements. In the present work, focusing on the noiseless case, we aim to disentangle the two distinct sub-tasks entailed in the Phase retrieval problem, namely the hard combinatorial problem of retrieving the missing signs of the measurements, and the nested convex problem of regressing the input-output observations to recover the hidden signal. To this end, we introduce and analytically characterize a two-level formulation of the problem, that we call ``Phase selection''.
Within the Replica Theory framework, we perform a large deviation analysis to characterize the minimum mean squared error achievable with different guesses for the hidden signs. 
Moreover, we study the free-energy landscape of the problem when both levels are optimized simultaneously, as a function of the dataset size. At low temperatures, in proximity to the Bayes-optimal threshold --- previously derived in the context of Phase retrieval --- we detect the coexistence of two free-energy branches, one connected to the random initialization condition and a second to the signal. We derive the phase diagram for a first-order transition after which the two branches merge. 
Interestingly, introducing an $L_2$ regularization in the regression sub-task can anticipate the transition to lower dataset sizes, at the cost of a systematic bias in the signal reconstructions which can be removed by annealing the regularization intensity.
Finally, we study the inference performance of three standard meta-heuristics in the context of Phase selection: Simulated Annealing, Approximate Message Passing, and Langevin Dynamics on the continuous relaxation of the sign variables. With simultaneous annealing of the temperature and the $L_2$ regularization, the algorithms are shown to approach the Bayes-optimal sample efficiency.
\end{abstract}

\maketitle


\section{Introduction}
\label{sec:intro}

Phase retrieval is a classical and pervasive problem in various scientific fields, originating in the 1950s within physics, expanding across disciplines such as X-ray crystallography, astronomy, and optics \cite{pr_optics,Millane:90}, and appearing even in non-imaging contexts including computer-generated holography, optical computing, acoustics, and quantum mechanics \cite{pr_accoustics,CORBETT200653}. The fundamental challenge lies in reconstructing a signal from the magnitude of a set of linear measurements while lacking phase information, resulting in a non-convex optimization landscape where conventional algorithms often fail to achieve good sample efficiency.

In recent years, Phase retrieval has also received much attention in the theoretical computer science community. Its real-valued version constitutes one of the simplest instances of non-convex continuous optimization, often studied as a tractable neural network prototype with non-linear activation functions \cite{Saglietti_2020, Mannelli_PR, mondelli_fundamental_2019}. In the analogy, the measurement vectors correspond to input samples, the hidden signal represents the optimal configuration of network weights, and the measured magnitudes are analogous to output labels. This interpretation provides a fertile ground for exploring key themes in learning theory, such as optimization in non-convex landscapes, the impact of initialization and optimization heuristics on generalization, and the role of training dataset size \cite{MignaccoStoch, EscapingMediocrity,Mignacco_2021}. 

From a statistical physics perspective, phase retrieval is an example of high-dimensional signal estimation, characterized by an intricate energy-entropy competition where the true signal is concealed within a vast landscape rugged with local minima. Such "crystal hunting" problems ---as they are informally denoted in the physics of disordered systems \cite{krzakala_statistical_2015}--- exhibit an algorithmically hard phase where all known polynomial-time algorithms struggle, becoming trapped in local minima with low overlap with the true signal \cite{Ricci_Tersenghi_2019, PhysRevX.9.011020}. 
An interesting observation, however, is that from the information-theoretical point of view, Phase retrieval becomes possible ---in principle--- as soon as the number of measurements equals the dimensionality of the signal, as in a simple linear system \cite{doi:10.1073/pnas.1802705116, BANDEIRA2014106, Saglietti_2020}. Yet the loss of phase information strongly hinders the inference performance of different local-search algorithms, making the problem difficult to solve efficiently without additional constraints or prior knowledge. For example, gradient descent was shown to struggle in the signal retrieval task \cite{Candes_2015} and therefore require good initialization ---e.g., through spectral methods \cite{candes2006robust, Candes_2015}--- or large amounts of data. Exploiting full knowledge of the data generative process, the Bayes-optimal approximate message-passing (AMP) approach to this problem was conjectured to achieve the best possible algorithmic performance in the large size limit. 

In this work, we propose a new framing of the maximum-a-posteriori (MAP) setting of Phase retrieval, where one potentially deals with imperfect information on the signal prior. Inspired by \citep{obuchi2018statistical}, presenting a statistical mechanical analysis of Compressed sensing framed as a variable selection problem, we aim to investigate whether it is possible to isolate the hard combinatorial component of Phase retrieval, i.e. the correct selection of the missing phases --- or signs, in the real-valued setting---, and elucidate how strategic problem formulation can enhance both our understanding of the problem and the inference performance. Along this line, we introduce \emph{Phase selection}, a closely related two-level optimization problem, where the hidden signal is to be retrieved through a standard linear regression after a specific choice for the signs of the measurements. We provide a detailed statistical physics analysis of this framework, under standard simplifying assumptions about the data and signal distributions.
We also apply different meta-heuristics to this novel formulation, investigating the advantages of disentangling the two sub-tasks within Phase retrieval.

The outline of the work is the following: in Sec.~\ref{sec:setting}, we mathematically define Phase Retrieval and Phase Selection problems, providing the common properties of their solutions. In Sec.~\ref{AnaliticalInvstigation}, we present the large deviation and replica analysis for the system in the infinite system size limit, within the so-called proportional regime. In Sec.~\ref{sec:NumericalResults}, we present three different solvers for Phase Selection instances, based on Simulated Annealing, Approximate Message Passing and Langevin Dynamics, on the continuous relaxation of the signs, and we study their performances. The analytical and numerical findings together with their implications are discussed in Sec.~\ref{sec:discussion}.

\section{Problem Setting} \label{sec:setting}

\subsection{Phase retrieval}
\label{sec:PhaseRetrieval}

In the noiseless real-valued Phase retrieval problem we are given a measurement matrix $A\in \mathbb{R}^{M\times N}$ and a measurement vector $\mathbf{y}\in\mathbb{R}^M$, whose entries are assumed to be produced as:
\begin{equation}
y_i = \left|  \frac{\sum_{j} A_{ij} (x_0)_j}{\sqrt{N}} \right|,
\label{GeneratingProcessPR}
\end{equation}
where $\mathbf{x_0}\in\mathbb{R}^N$ is a hidden signal to be inferred from the observations and \( | \cdot | \) denotes the element-wise modulus operation. 

In our analysis, we are going to make the standard simplifying assumption that the measurement matrix $A$ has i.i.d. Gaussian entries, i.e. $A_{ij}\sim\mathcal{N}(0,1)$, and consider a factorized normal prior also for the hidden signal, i.e. $(x_0)_i\sim\mathcal{N}(0,1)$. Note that, it is possible to relax these irrealistic assumptions to more complicated distributions \citep{maillard2020phaseretrievalhighdimensions}, but introducing correlations leads to an unnecessarily complicated analysis for the purpose of this study. 
Moreover, we will focus on the so-called proportional scaling regime, where $N$ and $M$ grow to infinity at a fixed rate $\alpha = M/N$, with $\alpha > \alpha_{IT}=1$ \cite{BANDEIRA2014106}, i.e. a sufficiently large dataset for the signal to be information-theoretically detectable. 

In the maximum-a-posteriori setting, the signal recovery problem is mapped onto the optimization of a loss function: 
\begin{gather}
    \begin{aligned}
        \hat{\mathbf{x}} &= \arg\min_\mathbf{x} \mathcal{H}_{A,\mathbf{y}}(\mathbf{x}) \\  \mathcal{H}_{A,\mathbf{y}}(\mathbf{x}) &= \lVert \mathbf{y} - |A \mathbf{x}|
        \lVert_2^2 + \frac{\lambda}{2}\lVert \mathbf{x} \lVert_2^2,
        \label{OptimizationPhaseRetrieval}
    \end{aligned}
\end{gather}
which computes the Mean Squared Error (MSE) between the measurement vector and the output of the inferred model. The additional $L_2$ penalty ---of controllable intensity--- on the model weights, was shown to be helpful in recovering the signal in previous work \cite{ma2018approximate}.

Note that often one considers a formally equivalent, but slightly less nefarious, formulation where the loss entails the MSE between the squared measurements of true and inferred signals, avoiding the singular absolute value operation and simplifying the landscape for gradient-based optimization. However, in this work, we are interested in maintaining a straightforward connection to linear regression, as it will become clearer in the next paragraph.

\subsection{Phase selection problem}

The core idea of the Phase selection formulation is to resolve the degeneracy introduced by the modulus operation, and explicit the signs ---i.e., real phases--- to be associated with the measurements via an auxiliary binary variable $\mathbf{S}\in\{-1,1\}^M$. Then, the loss function Eq.\eqref{OptimizationPhaseRetrieval} can be generalized to:
\begin{equation}
\mathcal{H}_{A,\mathbf{y}}(\mathbf{x}, \mathbf{S}) = \lVert \mathbf{S} \odot \mathbf{y} - A \mathbf{x}
 \lVert_2^2 + \frac{\lambda}{2}\lVert \mathbf{x} \lVert_2^2,
\label{EnergyFunction}
\end{equation}
where $\odot$ denotes the element-wise multiplication. Thus, the original signal estimation problem is decomposed into two optimization levels, the combinatorial problem of choosing the phases, $\mathbf{S}$, and the nested convex problem of estimating the signal, $\mathbf{x}$. For each sign selection, the \emph{internal} problem reduces to a simple linear regression, solved in closed form by the Ridge estimator:
\begin{equation}
\hat{\mathbf{x}}(\mathbf{S}|\mathbf{y},A) = \arg\min_{\mathbf{x}} \mathcal{H}_{A,\mathbf{y}}(\mathbf{x}, \mathbf{S}) = \left( A^\top A + \lambda \mathbb{I} \right)^{-1} A^\top \mathbf{S} \odot \mathbf{y}.
\label{eq:OptimizationPhaseSelection}
\end{equation}
On the other hand, the loss achieved in correspondence of this minimum induces a probability distribution over the sign configurations.

When the regularization is completely switched off, the unique global minimum of the generalized loss function Eq.~\eqref{OptimizationPhaseRetrieval} still achieves exactly $0$ mean squared error, with a perfect sign selection and retrieval of the hidden signal. The presence of a non-zero $L_2$-regularization could bias $MSE_x$ by explicitly reducing the norm of the inferred signal.
Remarkably, the complete structure of local minima of Phase retrieval is preserved, since each minimum in the original problem is in one-to-one correspondence with a specific gauge for the signs ---made explicit by removing the modulus operation after local convergence. However, note that the Phase selection formulation can introduce new local minima compared to Phase retrieval, since in some cases the Ridge estimator might not be able to yield a configuration that matches all the selected signs $\mathbf{S}$. Since Phase selection only introduces additional sub-optimal local minima, the information-theoretic threshold for exact recovery remains unchanged from that of the Phase retrieval problem \cite{BANDEIRA2014106}.

Given a locally optimal choice of signs $\mathbf{\hat{S}}$, stable against single spin flips, the two natural measures of the inference performance are given by: 

\begin{gather}
    \begin{aligned}
        MSE_y &= \frac{1}{M}\left\lVert \mathbf{\hat{S}}\odot\mathbf{y} - \frac{A ~\hat{\mathbf{x}}(\mathbf{\hat{S}}|\mathbf{y},A)}{\sqrt{N}}\right\rVert^2_2 \\
    MSE_x &= \frac{1}{N}\left\lVert \mathbf{x_0}-\hat{\mathbf{x}}(\mathbf{\hat{S}}|\mathbf{y},A) \right\rVert^2_2,
    \label{eq:energies}
    \end{aligned}
\end{gather}

respectively representing the mean squared error between the measurements, and the mean squared error associated with the hidden signal reconstruction.
Closely connected to those metrics, one can also measure the two overlaps:

\begin{gather}
    \begin{aligned}
    m &= \frac{\mathbf{x_0} \cdot \mathbf{\hat{x}} (\mathbf{\hat{S}}|\mathbf{y},A)}{N}\\
    O &= \frac{\mathrm{sign}\left(A ~\mathbf{x}_0\right) \cdot \mathbf{\hat{S}}}{M},
    \end{aligned}
\end{gather}

which should approach unity in the limit of perfect recovery.
The goal of our theoretical analysis, presented in Sec.\ref{AnaliticalInvstigation}, is to obtain an exact estimate of the inference performance as a function of the above-defined overlaps, making use of the Replica method \cite{Replica,Replica2}.

\subsubsection{Probabilistic formulation  of Phase Selection}

It is possible to reformulate Phase Retrieval problem from a Bayesian perspective: the estimate $\mathbf{x}$ is drawn from a posterior measure $p(\mathbf{x}|A,\mathbf{y})$ where the missing phases do not appear explicitly. Using the marginal probability theorem, we can explicit the dependency on $\mathbf{S}$, rewriting the posterior probability for $\mathbf{x}$ as 

\begin{align}
    p(\mathbf{x}|A,\mathbf{y}) = \int_{\Omega_S} p(\mathbf{x}|A,\mathbf{y},\mathbf{S}) p(\mathbf{S}|A,\mathbf{y}) d\mathbf{S}
    \label{eq:prob}
\end{align}
where $\Omega_S$ is the set of all possible sign assignments, $p(\mathbf{S}|A,\mathbf{y})$ is the probability of selecting $\mathbf{S}$ given the dataset $(\mathbf{y},A)$, and $p(\mathbf{x}|A,\mathbf{y},\mathbf{S})$ is the likelihood of  $\mathbf{x}$ given a realization of the signs $\mathbf{S}$. From Eq.\eqref{eq:prob}, the two-level optimization structure of Phase Retrieval becomes clear: we first infer the missing phases, and then we use them to obtain an estimate for the model.





\section{Analytical investigations}
\label{AnaliticalInvstigation}

\subsection{Large deviation analysis} \label{sec:large deviation}

We perform a large deviation analysis to describe the available energy levels ---denoted with $e$ in the following--- for the local minima of the loss in Eq.~\eqref{EnergyFunction}, as a function of the different sign selections. As customary in analogous high-dimensional settings \cite{PartialAnnealing}, we assume the local minima probability distribution obeys a large deviation principle: for a given measurement ratio $M/N = \alpha = \mathcal{O}(1)$, and assuming a fixed fraction of correct signs $O$, exponentially many sign configurations will yield an energy value $e$. The multiplicity of these equivalent configurations, $e^{N\Sigma(O,e)}$, is controlled by the rate function $\Sigma(O,e)$ ---called complexity in statistical physics. 
The large deviation principle implies that a random sample of the signs, at overlap $O$ with the correct assignment, will induce with high probability an energy close to the \emph{typical} value of the energy $\bar{e}(O)=\arg \max_{e} \Sigma(O,e)$. Lower-lying local minima ---sought in the Phase selection problem--- are instead exponentially rare \cite{Cui2020LargeDF}. 
Thus, studying the complexity function can aid the interpretation of the performance of local-search algorithms that try to bias the optimization towards these rare samples.

For technical reasons, in the following, we will access the complexity via the computation of its Legendre transform \citep{Cui2020LargeDF}, the so-called free entropy $\Phi(\mu, \phi)$, where $\mu$ is an inverse temperature biasing towards lower energy levels, and $\phi$ is a chemical potential for correct sign choices. In the thermodynamic limit $N \to \infty$, the saddle-point method yields:
\begin{align}
    \Phi(\mu, \phi) = \text{extr}_{e,O} \{\Sigma(O, e) + \mu \, e + \phi \, O\},
\end{align}
where $e = \partial_\mu \Phi(\mu, \phi)$ and $O = \partial_\phi \Phi(\mu, \phi)$. This relation implies that, for any value of $\mu$ and $\phi$, an interplay of entropy and energy causes the probability measure to concentrate on local minima with specific energy and sign overlap, which can be obtained by extremizing the free entropy. Then, by inverting the Legendre transform one can recover the sought complexity:
\begin{align}
    \Sigma(O, e) = \left[\Phi(\mu, \phi) - \mu \, e -\phi \, O \right]_{\partial_\mu\Phi= e, \partial_\phi\Phi=O}.
\end{align}



In the next section, we will present a concise summary of our analytical computation, employing well-established techniques from the physics of disordered systems \cite{mezard2009information}.


\subsection{Replica Symmetric formula for the large deviations}

To compute the quenched free-entropy for Phase selection, the starting point is the definition of the partition function $\Xi$, associated with the probability distribution over the sign configurations. Introducing an inverse temperature $\mu$ and a chemical potential $\phi$, we can write:
\begin{align}
    \Xi\left(\mu, \phi\,|\,A,\mathbf{x_0}\right)=\sum_{\left\{ S\right\} }e^{\phi\sum_{\nu=1}^M S_{\nu}-\mu \mathcal{H} \left(\mathbf{S}\,|\,A,\mathbf{x_0}\right)} \label{eq:partition_func}
\end{align}
where the energy of the internal system, $\mathcal{H} \left(\mathbf{S}\,|\,A,\mathbf{x_0}\right)$, is recovered as the maximum-a-posteriori, i.e. in the zero temperature limit of the posterior, for the estimated signal:
\begin{align}
\mathcal{H}\left(\mathbf{S}\,|\,A,\mathbf{x_0}\right)=\lim_{\rho\to\infty}-\frac{1}{\rho}\log\int \prod_{i=1}^Ndx_i P\left(x_{i}\right)\frac{\exp\left[-\rho \left(\sum_{\nu=1}^M \left(\mathbf{S}_{\nu}\frac{\boldsymbol{A^{\nu}}\cdot \mathbf{x_0}}{\sqrt{N}}-\frac{\boldsymbol{A^{\nu}}\cdot \mathbf{x}}{\sqrt{N}}\right)^{2}-\frac{\lambda}{2}\lVert \mathbf{x} \lVert_2^2.\right)\right]}{2\pi/\rho}.
\label{eq:MAP}
\end{align}
and where $A^\nu$ represents a row of the measurement matrix. Note that, for convenience of notation, compared to the definition Eq.~\eqref{eq:energies} for the $MSE_y$, we have considered the gauge transformation:
\begin{equation}
    \mathbf{S}\odot\left|\frac{A \mathbf{x}_0}{\sqrt{N}}\right| \to \mathbf{S}\odot\frac{A \mathbf{x}_0}{\sqrt{N}},
    \label{eq:gaugeTransform}
\end{equation}
which corresponds to reabsorbing in $\mathbf{y}$ the correct signs for the measurements. This reparametrization does not affect the evaluation of the complexity function but turns out to make the final expression easily interpretable: $S_\nu=1$ now corresponds to a correct sign selection for the corresponding measurement, while $S_\nu=-1$ corresponds to an incorrect one. This also allows the simplified expression for the chemical potential term in Eq.~\eqref{eq:partition_func}: a large $\phi$ pushes towards positive --- and thus correct --- signs.

In this type of analysis, one aims to trace out the dependence on specific realizations of the dataset --the quenched disorder-- and compute the average of the log-partition function density, i.e. the free entropy:
\begin{align*}
    \Phi\left(\mu, \phi\right)=\lim_{N\to\infty} \frac{1}{N}\mathbb{E}_{A,\mathbf{x_0}}\ln \Xi\left(\mu, \phi\,|\,A,\mathbf{x_0}\right).
\end{align*}
Following \cite{obuchi2018statistical}, we employ two separate instances of the replica trick to estimate this average. First, one can remove the intractable expectation of the logarithm, by using: 
\begin{align}
    \mathbb{E}_{A,\mathbf{x_0}}\ln  \Xi= \mathbb{E}_{A,\mathbf{x_0}}\frac{1}{s}\lim_{s\to 0} \Xi^s,
    \label{replica}
\end{align}
since the disorder average can be easily evaluated for integer values of $s$, and the result can be analytically continued to $s\to 0$. As for the second replica trick, we will assume the ratio between the two inverse temperatures in the problem --- external, $\mu$, and internal, $\rho$ --- to be integer, $\mu / \rho = \beta \in \mathbb{N}$ and then extrapolate the limit $\beta\to 0$. The replicated gran-canonical partition function can be rewritten as:
\begin{align*}
    \Xi^{s}	&=	\sum_{\left\{ S^{a}\right\} }\prod_{a}e^{\phi\sum_{\nu}S_{\nu}^{a}}\left[\int\prod_{a=1}^{s}\prod_{c=1}^{\beta}d\mathbf{x}_{ac} P\left(\mathbf{x}_{ac}\right) \prod_{\nu=1}^{\alpha N}\frac{\exp\left(-\frac{\mu}{\beta}\sum_{ac}\left(S_{\nu}^{a}\frac{\boldsymbol{A^{\nu}}\cdot \mathbf{x_0}}{\sqrt{N}}-\frac{\boldsymbol{A^{\nu}}\cdot \mathbf{x}_{ac}}{\sqrt{N}}\right)^{2}-\sum_{ac}\frac{\lambda\mu}{2\beta}\lVert \mathbf{x_{ac}} \lVert_2^2.\right)}{\sqrt{2\pi/(\mu/\beta)}}\right],
\end{align*}
introducing two replica indices, $a\in \{1,..,s\}$ and $c\in\{1,...,\beta\}$.

After switching expectation and limit operations, one can recover the free entropy as:
\begin{align}
    \Phi\left(\mu, \phi\right)=\lim_{N\to\infty} \frac{1}{N}\frac{1}{s}\lim_{s\to 0}\lim_{\beta\to 0} \mathbb{E}_{A,\mathbf{x_0}}\Xi^s\left(\mu, \phi\,|\,A,\mathbf{x_0}, \beta\right).
    \label{eq:GenFreeEntropy}
\end{align}

\subsubsection{Definition of the Order parameters}

Once the disorder average in $\mathbb{E}_{A,\mathbf{x_0}}\Xi^s\left(\mu, \phi\,|\,A,\mathbf{x_0}, \beta\right)$ is performed, it is natural to identify the overlap order parameters of the model --- or summary statistics \cite{https://doi.org/10.1002/cpa.21422} ---, which represent a set of sufficient descriptors of the system in high dimensions: 
\begin{gather}
    \begin{aligned}
        \tilde{q} &= \frac{\mathbf{x_{0}}\cdot \mathbf{x_{0}}}{N},\\
        q_{ac,bd} &= \frac{\mathbf{x_{ac}}\cdot \mathbf{x_{bd}}}{N},\\
        m_{ac} &= \frac{\mathbf{x_{0}}\cdot \mathbf{x_{ac}}}{N},
        \label{OrderParameters}
    \end{aligned}
\end{gather}
where $\tilde{q}$ represents the norm of the hidden signal, $q_{ac,bd}$ the overlap between two signal estimates sampled from the above-defined measure, and $m_{ac}$ is the magnetization, i.e. the overlap between hidden and estimated signal. Through these order parameters, and the associated Lagrange multipliers (hereafter denoted with a hat symbol), one can express the free-entropy as a composition of an interaction, an entropic and an energetic part: 
\begin{align}  
\Phi\left(\tilde{q},q,m\right)=\lim_{s,\beta\to0} \frac{1}{s} [G_{i}+\ln G_{s}+\alpha\ln G_{E}],
\label{eq.FreeEnergy}
\end{align}
defined as:
\begin{gather*}
    \begin{aligned}
        G_{i}&=i\hat{\tilde{q}}\tilde{q}+i\sum_{\{a,b,c,d\} \in D}\hat{q}_{ac,bd}q_{ac,bd}+i\sum_{ac}\hat{m}_{ac}m_{ac},\\
        G_{S}&=\int dx_{0} P(x_0) \int\prod_{ac}d x_{ac}P\left(x_{ac}\right)e^{-\frac{\lambda \mu}{2\beta}} e^{-i\hat{\tilde{q}}\lVert x_{0}\rVert_2^{2}-i\sum_{\{a,b,c,d\} \in D}\hat{q}_{ac,bd} x_{ac} x_{bd}-i\sum_{ac}\hat{m}_{ac} x_{0} x_{ac}},\\
        G_{E}&=\sum_{\{S^{a}\} }e^{\phi\sum_{a}S^{a}}\int\frac{du_{0}d\hat{u}_{0}}{2\pi}\prod_{ac}\frac{du_{ac}d\hat{u}_{ac}}{2\pi}\times\\
        &\times e^{i\hat{u}_{0}u_{0}+\sum_{ac}i\hat{u}_{ac}u_{ac}-\frac{1}{2}\left(\left(\hat{u}_{0}\right)^{2}\tilde{q}+\sum_{ac,bd}\hat{u}_{ac}\hat{u}_{bd}q_{ac,bd}+2\hat{u}_{0}\sum_{ac}\hat{u}_{ac}m_{ac}\right)}\frac{e^{-\frac{\mu}{\beta}\sum_{ac}\left(S^{a}u_{0}-u_{ac}\right)^{2}}}{\sqrt{2\pi/(\mu/\beta)}},
    \end{aligned}
\end{gather*}
where $P(x_0)$ is the factorized prior for the hidden signal (that we assume normal), $P(x_{ac})$ is the factorized distribution of $x$ (that we assume flat) and $D = \{a,b,c,d|(a=b) \wedge (c\ge d)\} \vee \{a,b,c,d|(a>b) \wedge c,d\} $.

The $N\to\infty$ limit in Eq.\eqref{eq:GenFreeEntropy} invites a saddle-point approximation of the integral over the possible values of the order parameters. Therefore, one needs to impose an extremum condition over the overlaps entering the expression of the free entropy. A parametric ansatz for the order parameters is introduced to restrict the search space for the extremum. 

\subsubsection{RS Ansatz}
\label{subsec:RSAnsatz}

We consider the simplest possible ansatz, the Replica Symmetric Ansatz \cite{Replica}, which assumes the order parameter matrices to be symmetric under the exchange of replica indices, implying: 
\begin{equation}
q^{ac,bd}=\begin{cases}
\begin{array}{c}
q_{2}\\
q_{1}\\
q_{0}
\end{array} & \begin{array}{c}
\text{{if} }a=b, \,\,c=d\\
\text{{if} }a=b,\,\,c>d\\
\text{{if} }a>b, \,\,\forall c, d
\end{array}\end{cases},
\end{equation}
and a replica-index-independent magnetization with the signal $m$.
Additionally, in the zero temperature limit $\rho\to\infty$ entailed in the MAP of Eq.~\ref{eq:MAP}, we impose the standard rescaling \cite{obuchi2018statistical} for the order parameters:
\begin{gather}
\begin{aligned}
    q_{2}=q_{1}+\frac{\delta q}{\rho};~~~~\hat{q}_{1}-\hat{q}_{2}=\frac{\rho}{\mu}\delta\hat{q};\\
    \hat{q}_{1/0}=\left(\frac{\rho}{\mu}\right)^{2}\hat{q}_{1/0};~~~~\hat{m}=\frac{\rho}{\mu}\hat{m}.
\end{aligned}
\end{gather}

The saddle-point values of $\hat{\tilde{q}}$ and $\tilde{q}$ can be determined independently of the other order parameters, since one can require $\phi(s=0)=0$. In this setting, we have $\tilde{q}=\int dx_{0}P(x_{0})\,x_{0}^{2}$ and $\hat{\tilde{q}}=0$, and with a Gaussian prior for the true signal $\tilde{q}=\int\mathcal{D}x_{0}\,x_{0}^{2}=1$. Since the saddle point value for $\hat{\tilde{q}}=0$, we have the following expression for the interaction term in the free entropy:
\begin{align}
    g_i = \lim_{\beta\to\infty} \lim_{s\to0}\frac{G_{i}}{s}=\frac{1}{2}\left(\delta\hat{q}q_{1}-\frac{\hat{q}_{1}\delta q}{\mu}-\hat{q}_{1}q_{1}+\hat{q}_{0}q_{0}-2\hat{m}m\right).
\end{align}
The final expression for the free entropy, after the $s\to 0$ limit is simply $\Phi = g_i + g_S + \alpha g_E$ , where the entropic and energetic terms read:
\begin{align}
    g_S = \lim_{\beta\to\infty}\lim_{s\to0}\frac{\ln G_{S}}{s}= - \frac{1}{2}\log\left(\frac{(\delta\hat{q}+\lambda)-\left({\hat{q}_{1}-\hat{q}_{0}}\right)}{(\delta\hat{q}+\lambda)}\right)+\frac{\hat{q}_{0}+\hat{m}^{2}}{2\left((\delta\hat{q}+\lambda)-\left(\hat{q}_{1}-\hat{q}_{0}\right)\right)}, \label{eq:entropic}
\end{align}
\begin{gather}
\begin{aligned}
    g_E = \lim_{\beta\to\infty}\lim_{s\to0} \frac{\ln G_{E}}{s}&=\frac{1}{2}\log\left(\frac{1+2\delta q}{1+2\delta q+2\mu\left(q_{1}-q_{0}\right)}\right) -\mu\frac{\tilde{q}+q_{0}}{\left(1+2\delta q\right)+2\mu\left(q_{1}-q_{0}\right)}\\
    &+\int\mathcal{D}z_{0}\int\mathcal{D}u_{0}\log\left[2\cosh\left(\phi+2\mu\frac{\sqrt{\tilde{q}}\left(mu_{0}+\sqrt{q_{0}-m^{2}}z_{0}\right)u_{0}}{\left(1+2\delta q\right)+2\mu\left(q_{1}-q_{0}\right)}\right)\right], \label{eq:energetic}
\end{aligned}
\end{gather}
where we denoted a standard Gaussian measure with $\mathcal{D}x = \frac{e^{-x^2/2}}{\sqrt{2\pi}}dx$.
The saddle point equations can be obtained by imposing stationarity of the free entropy with respect to all the order parameters and their Lagrange multipliers. 

It is important to point out that the replica symmetric (RS) assumption, a common simplification in similar analyses \cite{Cui2020LargeDF}, may fail to capture the full complexity of the system. This occurs if the system transitions to a replica symmetry-broken (RSB) phase. Previous studies \cite{Cui2020LargeDF, doi:10.1073/pnas.1802705116} suggest that an RSB ansatz is likely unnecessary near the peak of the complexity curves, as confirmed by the local stability analysis provided in App.~\ref{subsec:stability}.
However, in more extreme regimes with large values of $\mu$, the system may enter a highly frustrated state where the complexity approaches zero, indicating that the system's phase space is sparsely populated with solutions and the RS assumption might be violated. 
Pursuing a complete RSB analysis in a large deviation context is technically challenging due to the largely increased computational cost of converging the saddle-point equations. In App.\ref{subsec:1RSB}, we provide the analytical form of these equations, but a thorough investigation of the RSB effects is left for future work. 

\subsection{Results} \label{sec:RSresults}


We can now seek the stationary values for the order parameters, and then use the inverse Legendre transform to recover the complexity function, $\Sigma$, for different settings of $\alpha, \mu, \phi, \lambda$. 

\subsubsection{$\lambda=0$, fixed sign overlap $O$}
We start by exploring the unregularized case, at an intermediate value $\alpha=1.7$. We recover the full large deviation of the complexity at fixed values of $O$, allowing us to study the impact of a strategic selection of the signs compared to random sampling. $O$ is fixed implicitly, by finding the conjugated value of the chemical potential $\phi$. 

\begin{figure}
    \centering
    \includegraphics[scale=0.3]{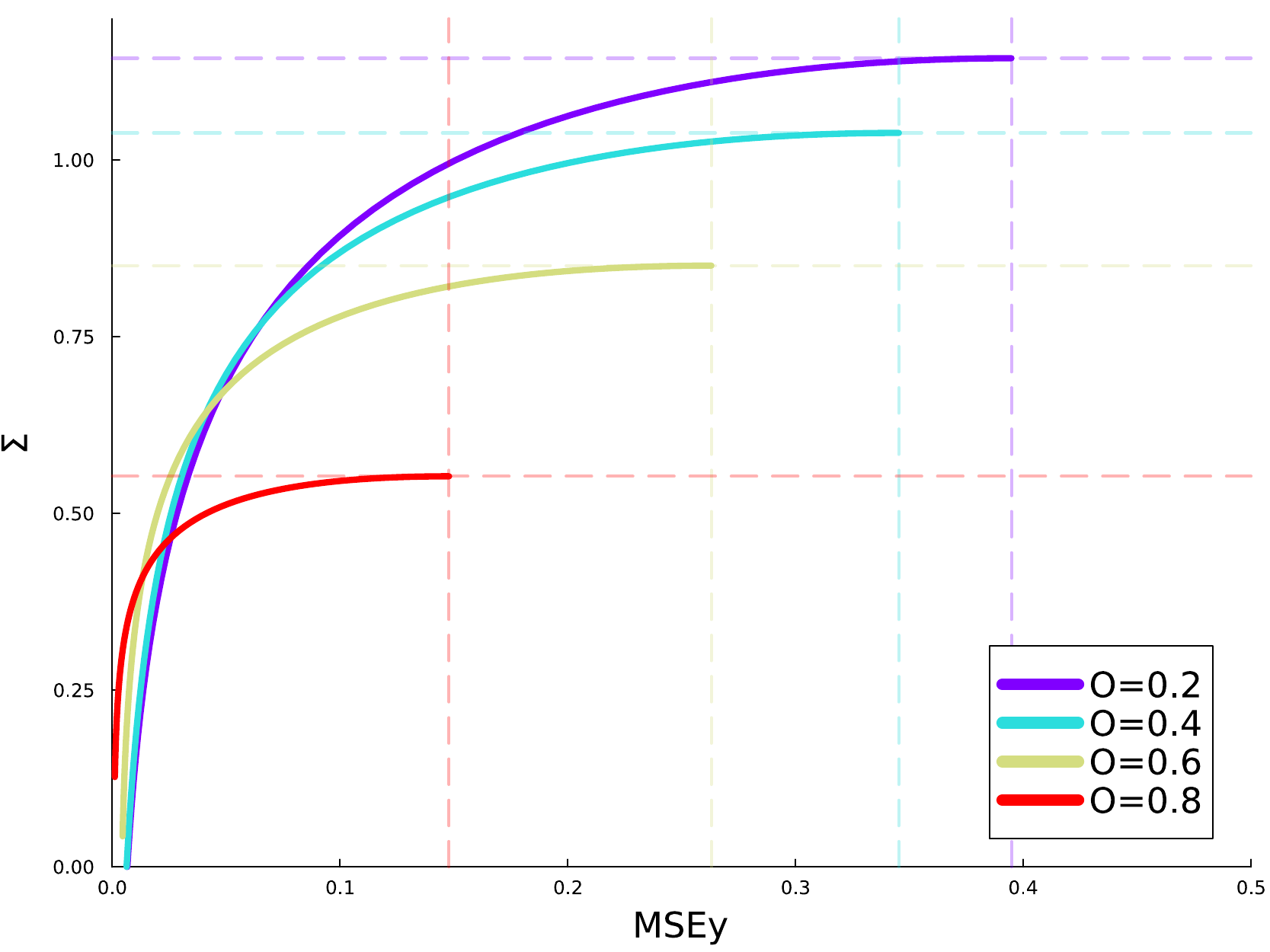}
    \includegraphics[scale=0.3]{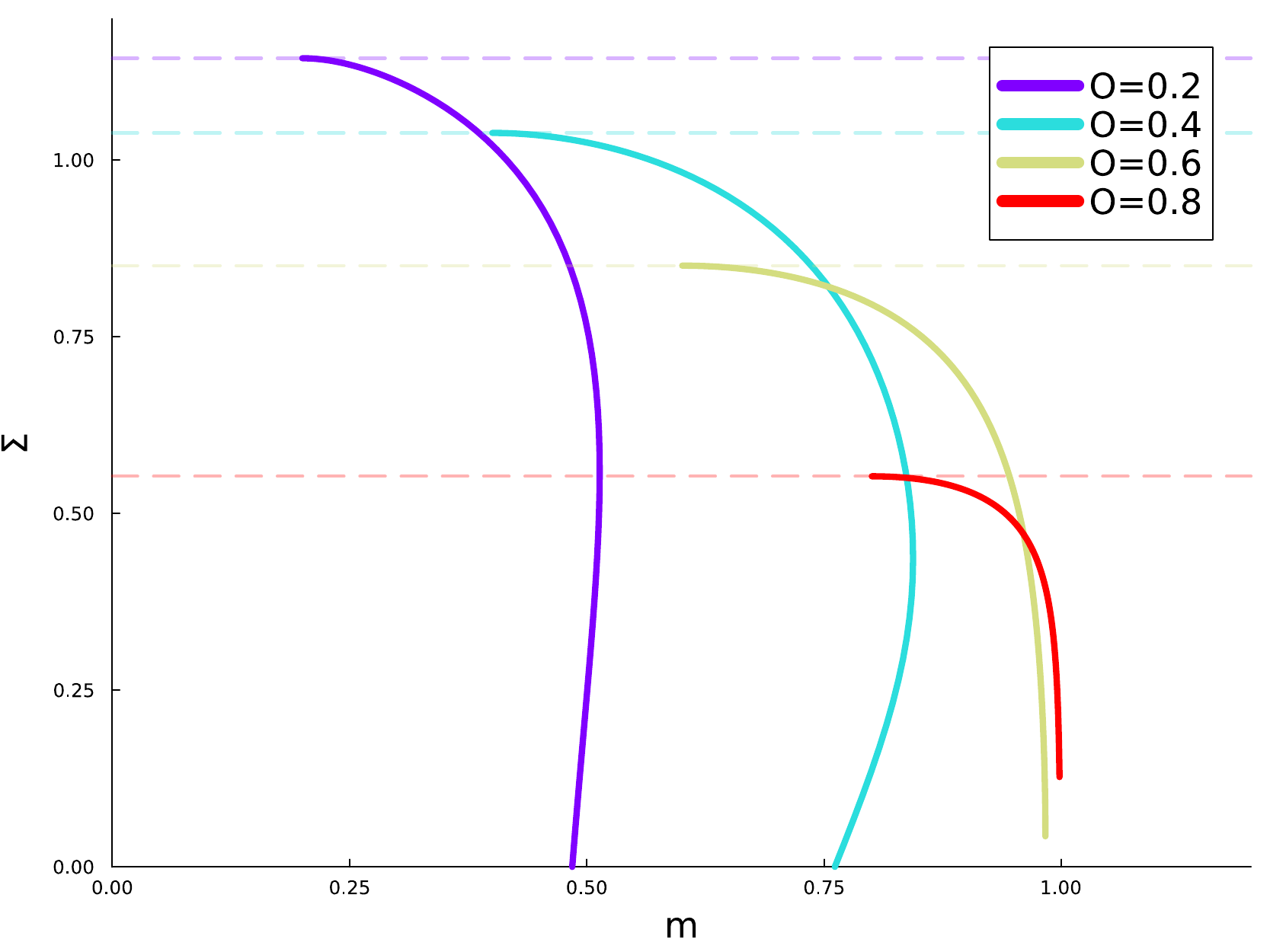}
    \caption{\textit{Left panel:} The complexity $\Sigma$ is represented as a function of the $MSEy$, i.e. the number of binary configurations having a fixed mean squared error on the labels. The reported curves have different values of the overlap among sign $O \in \{0.2, 0.4, 0.6, 0.8\}$, and are obtained with null regularization and fixed $\alpha=1.7$. The horizontal dashed line reports the expected maximum value for the complexity, given by the binomial distribution, while the vertical one outlines the largest $MSE_y$ value. \textit{Right panel:} The complexity $\Sigma$ is reported as a function of the overlap $m$, i.e. the number of binary configurations providing an estimate $\mathbf{x}$, having an overlap $m$ with $\mathbf{x_0}$. The parameters are the same used in the left panel. The horizontal dashed line reports the maximum complexity value.}
    \label{Fig:ComplexityVsMagnetization}
\end{figure}

The left panel of Fig.\ref{Fig:ComplexityVsMagnetization} illustrates the relationship between the complexity $\Sigma$ and the mean squared error $MSE_y$, for different values of the sign overlap, $O\in\{0.2,0.4,0.6,0.8\}$, $\alpha=1.7$. The curves are spanned by varying the inverse temperature $\mu$, i.e. low values of $\mu$ are associated to a more uniform choice of signs achieving higher complexity and higher $MSE_y$. As the overlap $O$ increases, the system's $MSE_y$ decreases. 
Notably, the maximum of the entropy curve, at $\mu=0$, achieves $\Sigma=-\alpha(\frac{1+O}{2}\log \frac{1+O}{2}+\frac{1-O}{2}\log\frac{1-O}{2})$ as one would expect from a completely random selection of the signs.

The right panel of Fig.~\ref{Fig:ComplexityVsMagnetization} displays the achieved $\mathbf{x}$ overlap with the signal, while the sign selections range from random to strategic. Interestingly, at low sign overlap $O$, the curves are characterized by a non-monotonic behavior of $m$ while $\Sigma$ decreases, which indicates that the loss is not necessarily minimized when the magnetization is maximum.

\subsubsection{$\lambda=0$, $\phi=0$ (unconstrained $O$)}

In an optimization scenario, the correct phases are unknown, making it unfeasible to bias toward a correct sign selection with the potential $\phi$. We thus set $\phi=0$ in the following, allowing the overlap $O$ to vary with $\mu$. To assess the recoverability of the hidden signal, one can study the free entropy $\Phi$ as a function of the inverse temperature. 

At large values of $\mu$ and sufficiently small $\alpha$, two distinct branches of free entropy coexist: the \emph{uninformed} branch, characterized by low magnetization with the true signal, $m\sim0$, and the \emph{informed} branch, reaching signal recovery $m\to1$ in the limit $\mu\to\infty$. The informed branch appears above the so-called spinodal transition $\mu_{SP}$, where the free entropy develops a maximum in correspondence of the signal, and then becomes the dominant branch at $\mu_{F}$, where said maximum becomes the global maximum of the free energy. 

The analysis implies that a $\mu$-annealing procedure from a random (low magnetization) initialization will lead to retrieval of the true phases and signal only when $\alpha$ is sufficiently large, since the informed branch merges with the uninformed one in this setting. Instead, at lower values of $\alpha$ the two branches are detached and the $\mu$-annealing is bound to get trapped in the uninformed branch, preventing recovery.

\begin{figure}
    \centering
    \includegraphics[scale=0.3]{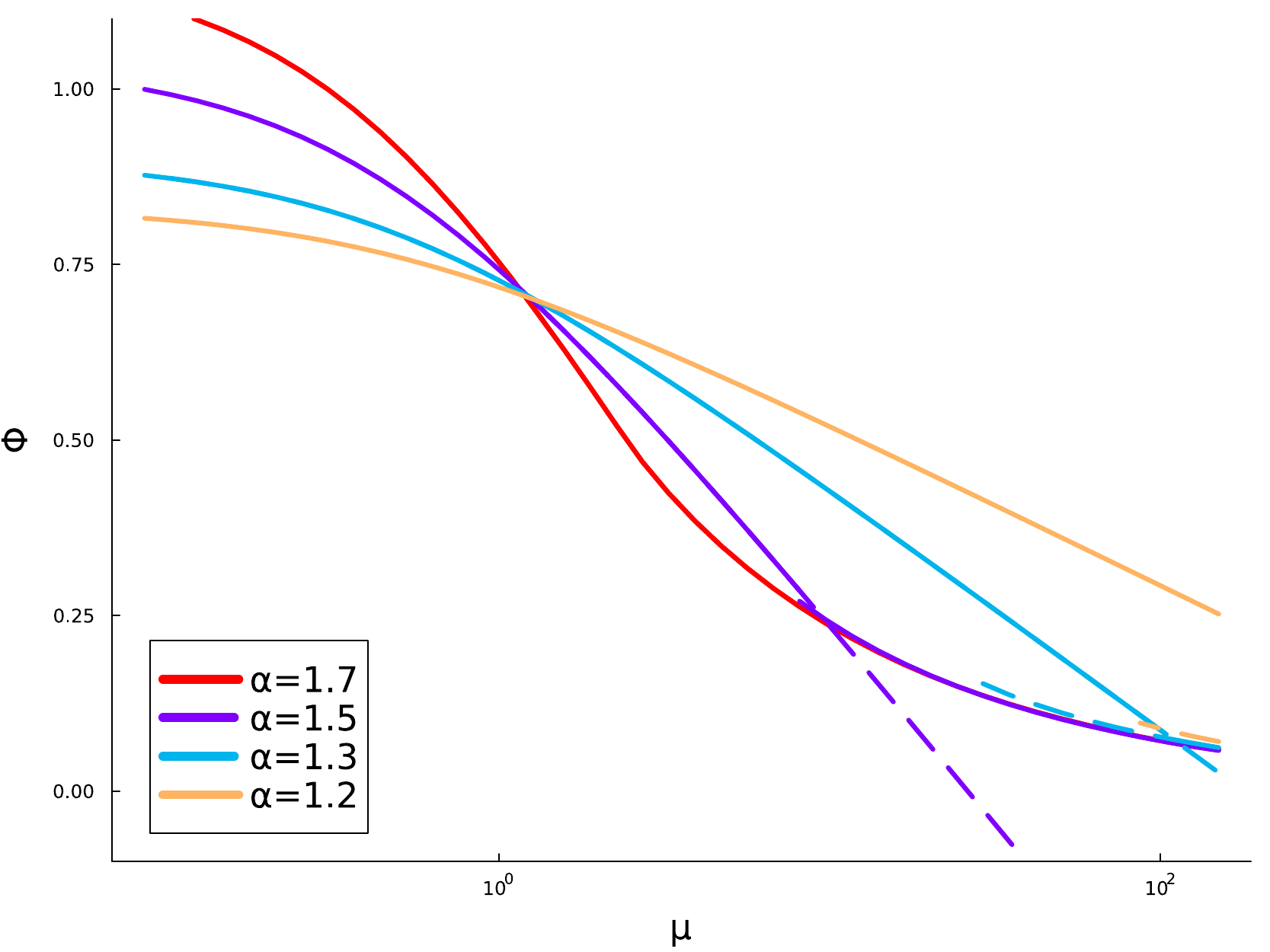}
    \includegraphics[scale=0.3]{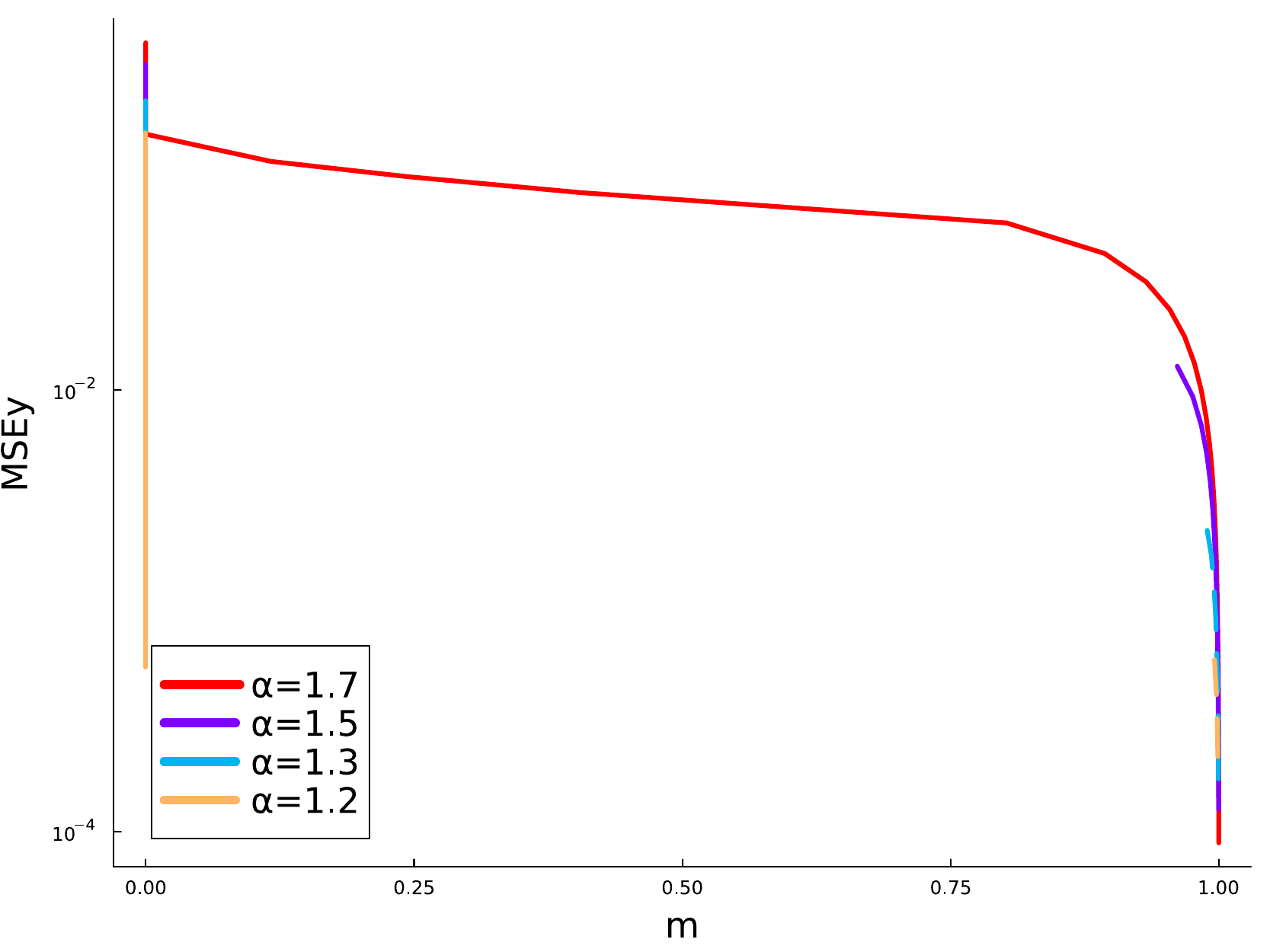}
    \caption{\textit{Left panel}: We report the Free Entropy $\Phi$ as a function of $\mu$, for various values of $\alpha\in\{1.2,1.3,1.4,1.5,1.6,1.7\}$ and null regularization. The informed branch and the uninformed one are reported. For each value of the inverse temperature $\mu$ the stable branch (the one having higher Free Entropy) is represented with a solid line, while the unstable one is reported with a dashed line. \textit{Right panel}: We report the mean square error on the labels $MSE_y$ as a function of $m$, for different value of $\alpha\in\{1.2,1.3,1.4,1.5,1.6,1.7\}$. The informed and uninformed branches are reported. At fixed inverse temperature the branch having the highest $\Phi$ is drawn with a solid line, the other with a dashed one.}
    \label{Fig:NoLambda}
\end{figure}

The left panel of Fig.~\ref{Fig:NoLambda} displays the coexisting free entropy branches (full lines for dominant branches, dashed line for sub-dominant ones), and shows that the merging happens for $\alpha = 1.7$. In the right panel of Fig.\ref{Fig:NoLambda}, one can see the corresponding behavior of the magnetization, plotted as a function of the fitting error $MSE_y$. 

In principle, the merging of the two branches in this analysis should signal a hard-easy algorithmic transition for Phase Selection. However, as we will describe in Sec.\ref{sec:NumericalResults}, we find a discrepancy between the predicted threshold and the algorithmic behavior of a simulated annealing-based solver, which can solve the problem only for higher values $\alpha>1.7$. We attribute this discrepancy to the presence of Replica symmetry-breaking effects, signaled by the local instability of the RS results at high values of $\mu$, shown in \ref{sec:RSStability}. However, we will also show that the qualitative insights provided by this incorrect analysis are still informative of the underlying properties of the optimization problem, and lead to the development of a novel efficient heuristic for solving Phase retrieval.

Note also that the predicted threshold $\alpha=1.7$ is still much larger than the Bayes-Optimal threshold of Phase retrieval, $\alpha_{BO}=1.13$.  
To bridge this gap, it is necessary to incorporate regularization, as we will discuss in the next paragraphs.

\subsubsection{Analysis of the stabilities, impact of $\mu$ and $\lambda$}
\label{subsec:Stabilities}

We have shown how annealing the inverse temperature $\mu$ may be sufficient to recover the hidden signal $\mathbf{x_0}$, provided $\alpha$ is large enough. We now turn to understanding exactly what happens to the signs' selection as $\mu$ increases, and whether switching on the regularization $\lambda$ can help to achieve a lower recovery threshold. Some insight about their impact can be derived from the analysis of the so-called \emph{stabilities}:
\begin{align*}
    \Delta^\nu = \frac{\mathbf{x}\cdot\boldsymbol{A}^{\nu}}{\sqrt{N}}, 
\end{align*}
representing the measurement amplitudes obtained with the signal estimate $\mathbf{x}$. In particular, for each stability, the inferred model also estimates a sign $S^\nu$, which needs to be compared to the missing phase produced by the true signal $\mathbf{x_0}$. To highlight the correlation between the amplitude of the stabilities and the correctness of the sign choices, as a function of $\mu$ and $\lambda$, we derive the analytical expression for the distribution of the stabilities $P(\Delta)$, and the distribution of the stabilities conditioned to a correct sign choice $P_*(\Delta)$:
\begin{align}
    P(\Delta) = \int \mathcal{D}z_0 \int \mathcal{D}u_0 \frac{\sum_{S\in\{-1,1\}}(1+2\delta q)\frac{e^{-\frac{z^*_+(\Delta, \delta_q, q_1, q_0, m, S)^2}{2}}}{\sqrt{2\pi(q_1-q_0)}}e^{-S\phi}e^{-\frac{\left(z^*_+(\Delta, \delta_q, q_1, q_0, m, S)+u_0\right)^2}{1+2\delta q}}}{\sqrt{\frac{1+2\delta q}{(1+2\delta q) + 2\mu (q_1-q_0)}} e^{-\mu\frac{\left(\sqrt{q_0-m^2}z_0 +m u_0\right) ^2 + u_0^2}{(1+2\delta q) + 2\mu (q_1-q_0)}}2\cosh\left( \phi + 2\mu \frac{\left(\sqrt{q_0-m^2}z_0 +m u_0\right) u_0}{(1+2\delta q) + 2\mu (q_1-q_0)} \right)},
    \label{eq:stabilities}
\end{align}
\begin{align}
    P_{*}(\Delta) = \int \mathcal{D}z_0 \int \mathcal{D}u_0 \frac{(1+2\delta q)\frac{e^{-\frac{z^*_+(\Delta, \delta_q, q_1, q_0, m, 1)^2}{2}}}{\sqrt{2\pi(q_1-q_0)}}e^{-\phi}e^{-\frac{\left(z^*_+(\Delta, \delta_q, q_1, q_0, m, 1)+u_0\right)^2}{1+2\delta q}}}{\sqrt{\frac{1+2\delta q}{(1+2\delta q) + 2\mu (q_1-q_0)}} e^{-\mu\frac{\left(\sqrt{q_0-m^2}z_0 +m u_0\right) ^2 + u_0^2}{(1+2\delta q) + 2\mu (q_1-q_0)}}2\cosh\left( \phi + 2\mu \frac{\left(\sqrt{q_0-m^2}z_0 +m u_0\right) u_0}{(1+2\delta q) + 2\mu (q_1-q_0)} \right)}.
\end{align}
where $z^*_+(\Delta, \delta_q, q_1, q_0, m, s) = \Delta\frac{1+2\delta q}{\sqrt{q_1-q_0}} - \frac{m u_0 + \sqrt{q_0-m^2}z_0 - s 2\delta q u_0}{\sqrt{q_1-q_0}}$. The expression of $P_*(\Delta)$ can be obtained from Eq.~\eqref{eq:stabilities} by fixing $S=1$. 
Details on the derivation of Eq.~\eqref{eq:stabilities} are reported in App.\ref{app:stabilities}.

\begin{figure}[t]
    \centering
    \includegraphics[scale=0.21]{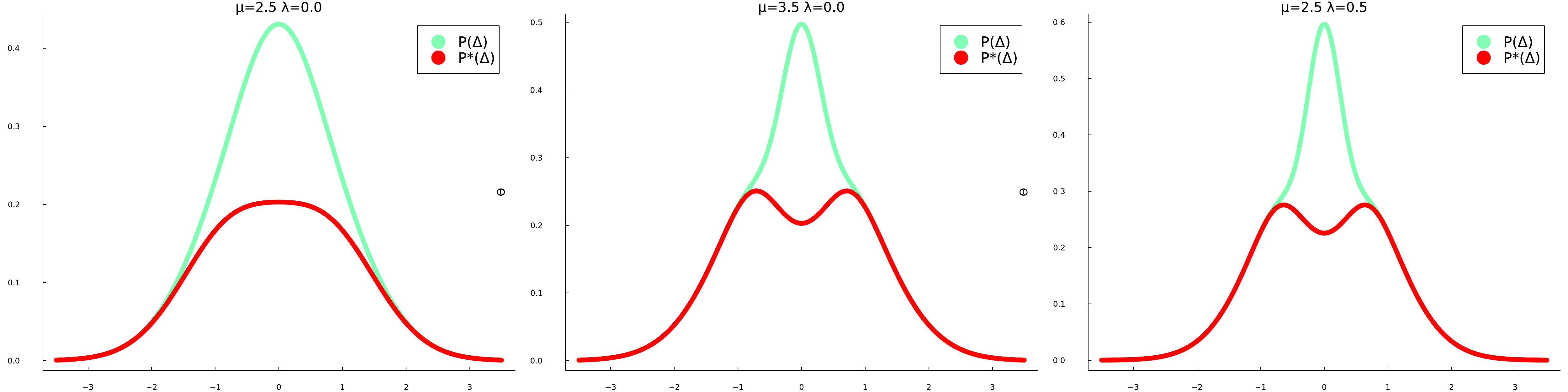}
    \caption{We present the istogram of stabilities for $O = 0.5$, $\alpha=1.7$, for three possible combinations of $\lambda$ and $\mu$ values. \textit{Left panel:} The inverse temperature is fixed to $\mu = 2.5$ and the regularization parameter to $\lambda=0.0$. \textit{Central panel:} The inverse temperature is $\mu = 3.5$ and the regularization parameter is $\lambda=0.0$. \textit{Right panel:} The inverse temperature is $\mu = 2.5$ and the regularization parameter is $\lambda=0.5$.} 
    \label{Fig:Stabilities} 
\end{figure}

In Fig.~\ref{Fig:Stabilities}, we display the distributions obtained with $\alpha=1.7$ and $O=0.5$, and different values of $\mu$ and $\lambda$. In particular, we show two perturbations of the same initial setting (left panel): increased $\mu$ (center), and increased $\lambda$ (right). 
Notably, the effect of increasing the inverse temperature mirrors that of increasing the regularization: both changes focus the mass of $P_*(\Delta)$ on stabilities with large magnitude, inducing a bimodal distribution, while the overall distribution $P(\Delta)$ squeezes around $\Delta=0$.
An intuitive explanation of this observation is that a wrong sign of a large magnitude stability has a larger impact on the signal reconstruction, as can be deduced from Eq.~\eqref{eq:OptimizationPhaseSelection}, and on the $MSE_y$, so correcting this type of errors first is the most efficient way of optimizing the loss. 


\subsubsection{The role of $\lambda$ in signal recovery}
\label{subsec:lambda}

From the results of Sec. \ref{subsec:Stabilities}, it is clear that the introduction of regularization can be beneficial since it has some similar effects to annealing $\mu$. Driven by this finding, we repeat the analysis of the $\mu$-annealing in the presence of a fixed $\lambda>0$. 

The top left panel of Fig.\ref{Fig:NoLambda} shows again the $MSE_y$ as a function of magnetization $m$ while $\mu$ is increased, this time with a regularization $\lambda=0.005$. By comparing the result with Fig.\ref{Fig:NoLambda}, it is evident that the merging of the branches happens above a smaller alpha threshold, allowing a random initialization to reach high magnetization for large values of $\mu$.
However, as the inverse temperature increases, the magnetization $m$ initially rises, reaches a peak, and can also decrease. This non-monotonic behavior is attributed to the biasing effect of regularization, which constrains the norm of the signal. A non-zero $\lambda$ prevents $\mathbf{x}$ from adequately aligning with $\mathbf{x_0}$, forcing $\mathbf{x}$ to lose the direction of the hidden signal. For $\alpha = 1.3$, we observe a range of $m$ in which both branches are unstable. This phenomenon is associated with the behavior of $\phi(\mu)$: as $\mu$ increases beyond $\mu_{Spinodal}$, the uninformed branch initially remains subdominant but eventually merges with the informed branch. 

\begin{figure}[h]
    \centering
    \includegraphics[scale=0.25]{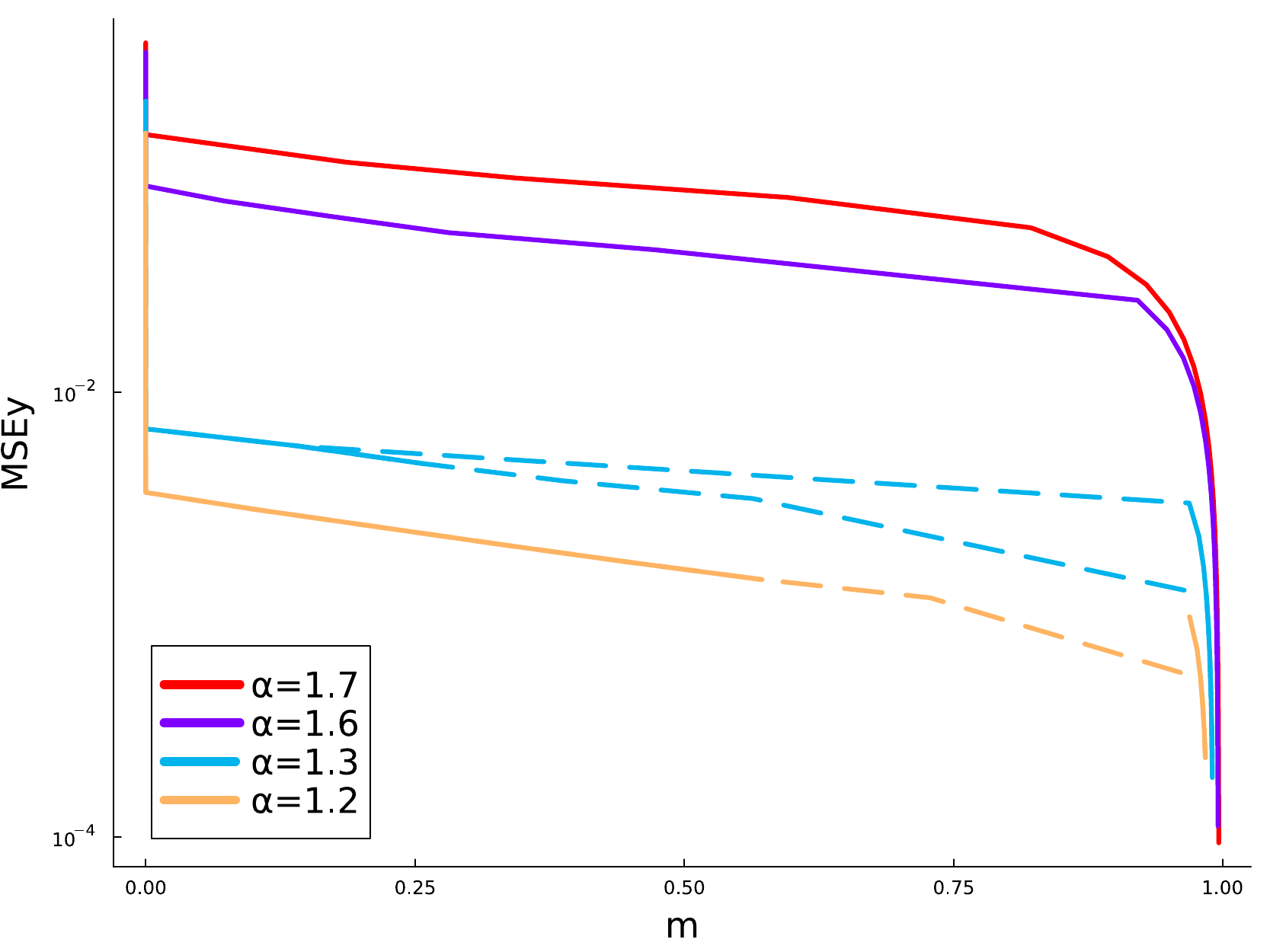}
    \includegraphics[scale=0.25]{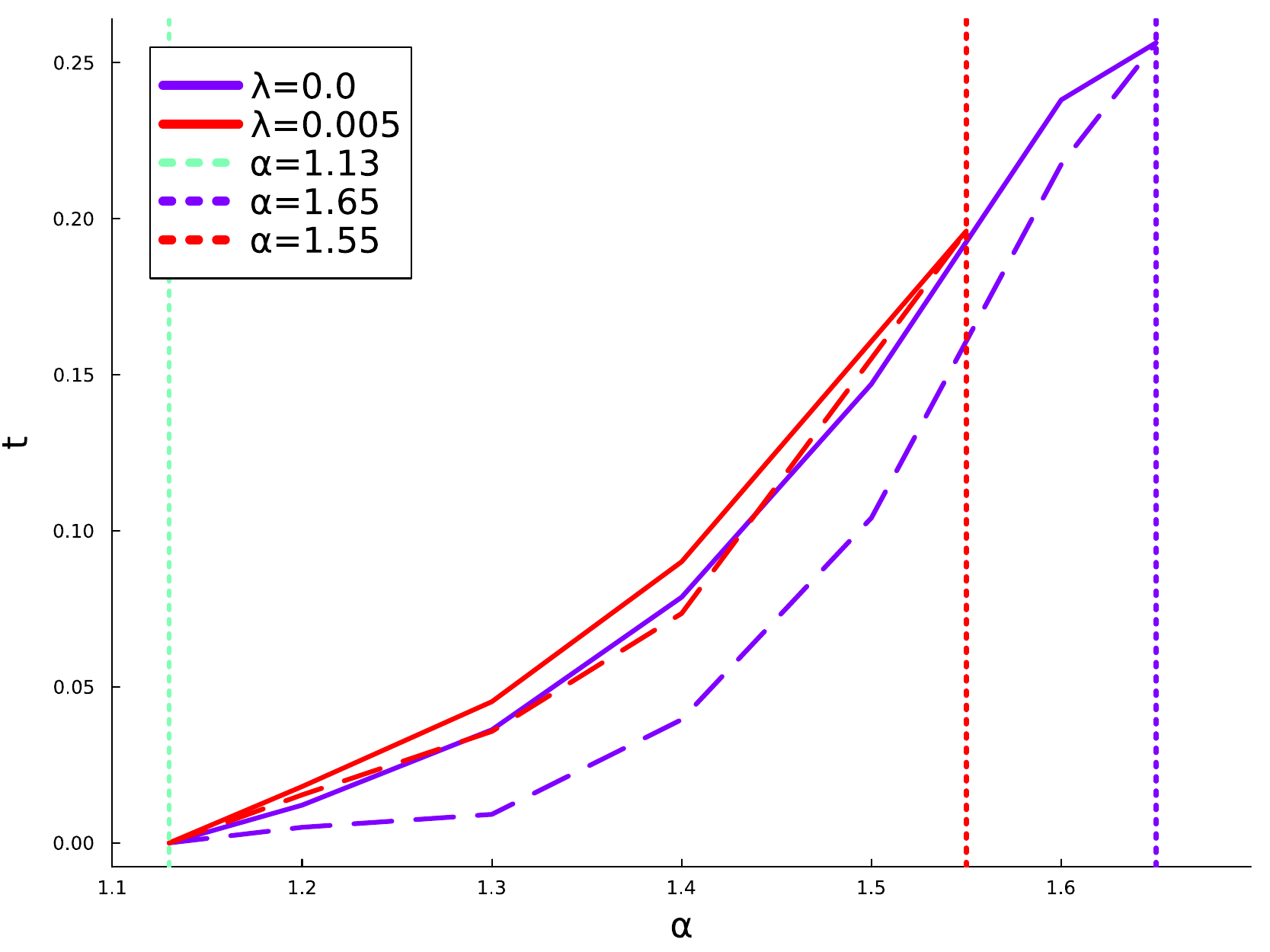}
    \includegraphics[scale=0.25]{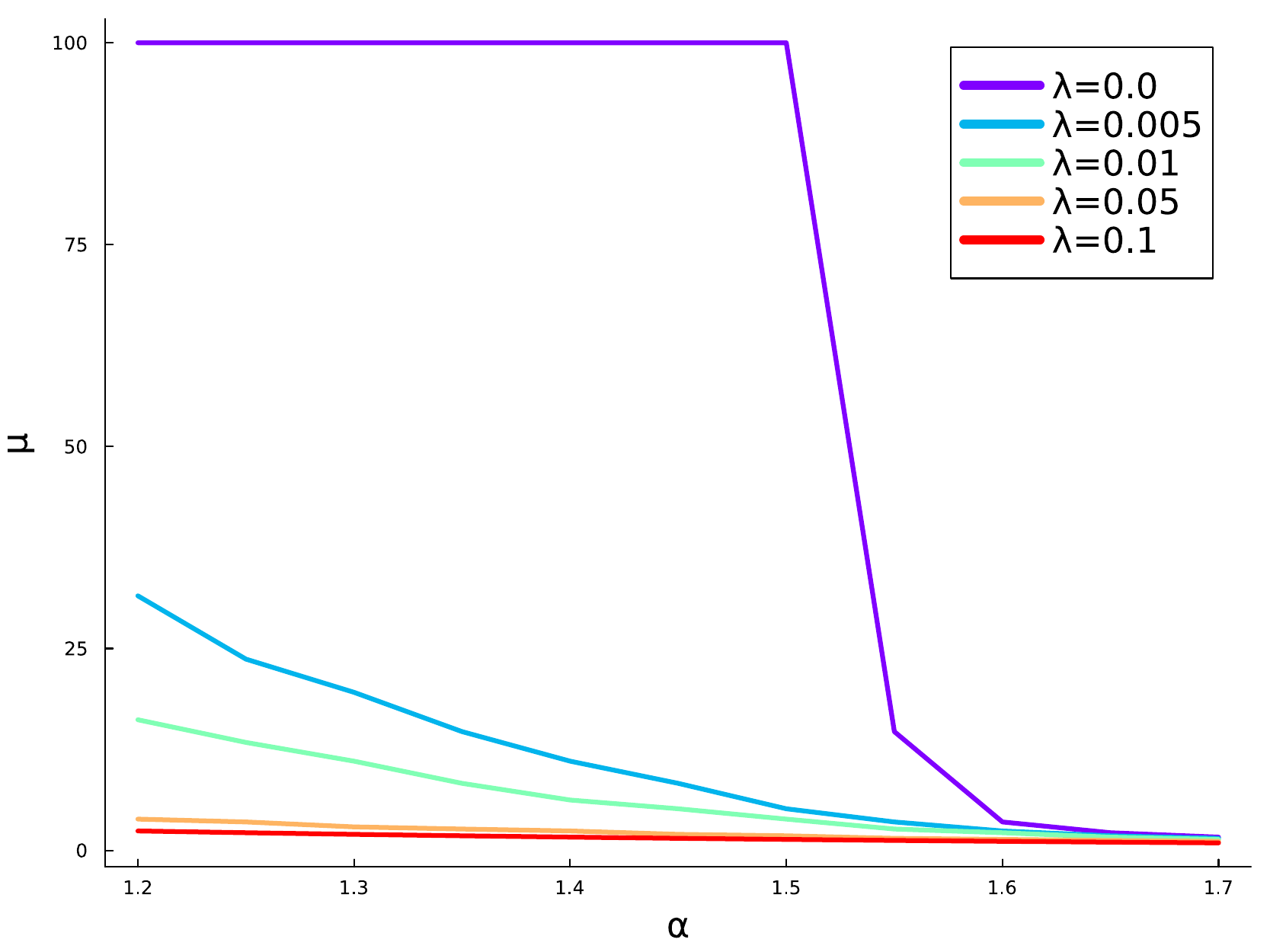}
     \includegraphics[scale=0.25]{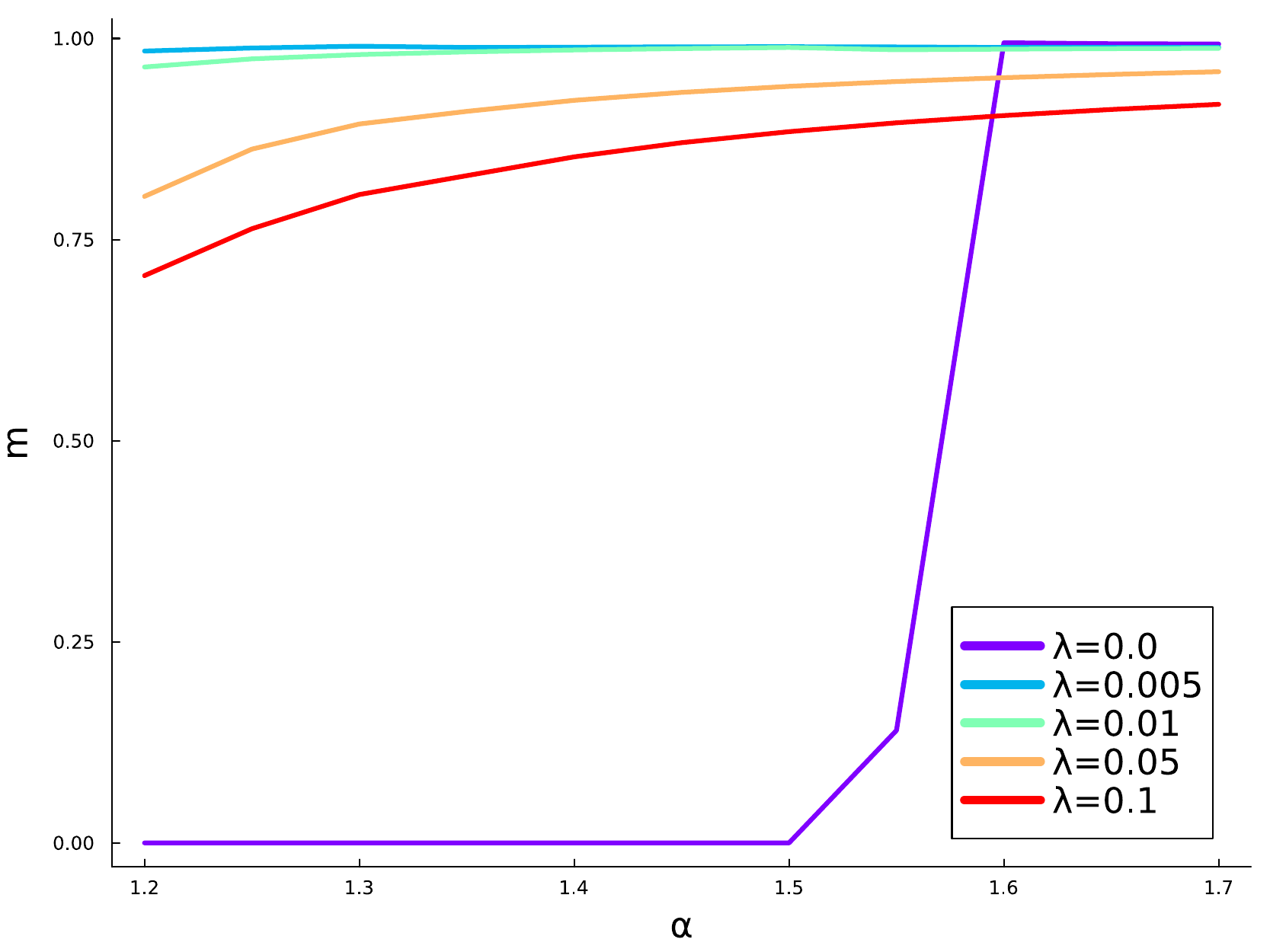}
    \caption{\textit{Top-left panel:} We present the mean square error on the labels $MSE_y$ vs $m$ for $\alpha\in\{1.2,1.3,1.4,1.5,1.6,1.7\}$, at null regularization. Both the informed and the uninformed branches are reported. The dashed lines are associated with the branch having lower free entropy at a specific value of $\mu$. \textit{Top-right panel:} The phase diagram is reported: for different values of $\alpha$ the temperature at which first-order transition (the dashed line) and spinodal transitions (the solid line) occur, for fixed $\lambda$. The red vertical line represents the information-theoretic threshold for the Phase Selection problem, and the purple and blue vertical lines represent the solvability threshold for the different values of $\lambda$. \textit{Bottom-left panel:} the plot presents the inverse temperature at which the uninformed branch reaches $m=0.1$ as a function of $\alpha$, for different values of regularization parameter. \textit{Bottom-right panel:} the plot the maximum $m$ obtained in the dynamics as a function of $\alpha$, for different values of $\lambda$, for the uninformed branch.}
    \label{Fig:Lambda}
\end{figure}

In the top right panel of Fig.\ref{Fig:Lambda}, we show the phase diagram of the problem and the effect of $L_2$ regularization. In particular, we identify the hard algorithmic region where the two free energy branches coexist and are detached, characterized by distinct values of $\mu$ for the spinodal and the first-order transitions. This hard window shrinks in the presence of regularization, which means that a $\lambda>0$ can be useful to escape the uninformed initialization trap. 

However, the bottom left and bottom right panels of Fig.\ref{Fig:Lambda} show that the regularization has a dual effect. On the one hand, on the left, we show that increasing $\lambda$ can anticipate the temperature at which the estimator starts to build up an overlap with the signal (e.g., in the figure we show the value of $\mu$ at which $m$ reaches $0.1$). On the other hand, the right panel shows that excessive regularization can impair complete signal reconstruction since the low norm bias will also reduce the maximum achieved overlap between $\mathbf{x}$ and $\mathbf{x_0}$. 
Note that $\mathbf{x_0}$ is the global minimum of the Phase Selection loss only for $\lambda = 0$.

\subsubsection{Simultaneous annealing of $\mu$ and $\lambda$}
\label{subsec:annealing}

The shown results suggest that, while the regularization can be beneficial at low inverse temperatures since it anticipates the alignment with the true signal, it can hinder the recovery process at larger values of $\mu$ where it can prevent full recovery.
This suggests the implementation of a
simultaneous annealing procedure, where the product $\mu \lambda$ is kept fixed as $\mu\to\infty$ and $\lambda\to0$. 


\begin{figure}[t]
    \centering
    \includegraphics[scale=0.3]{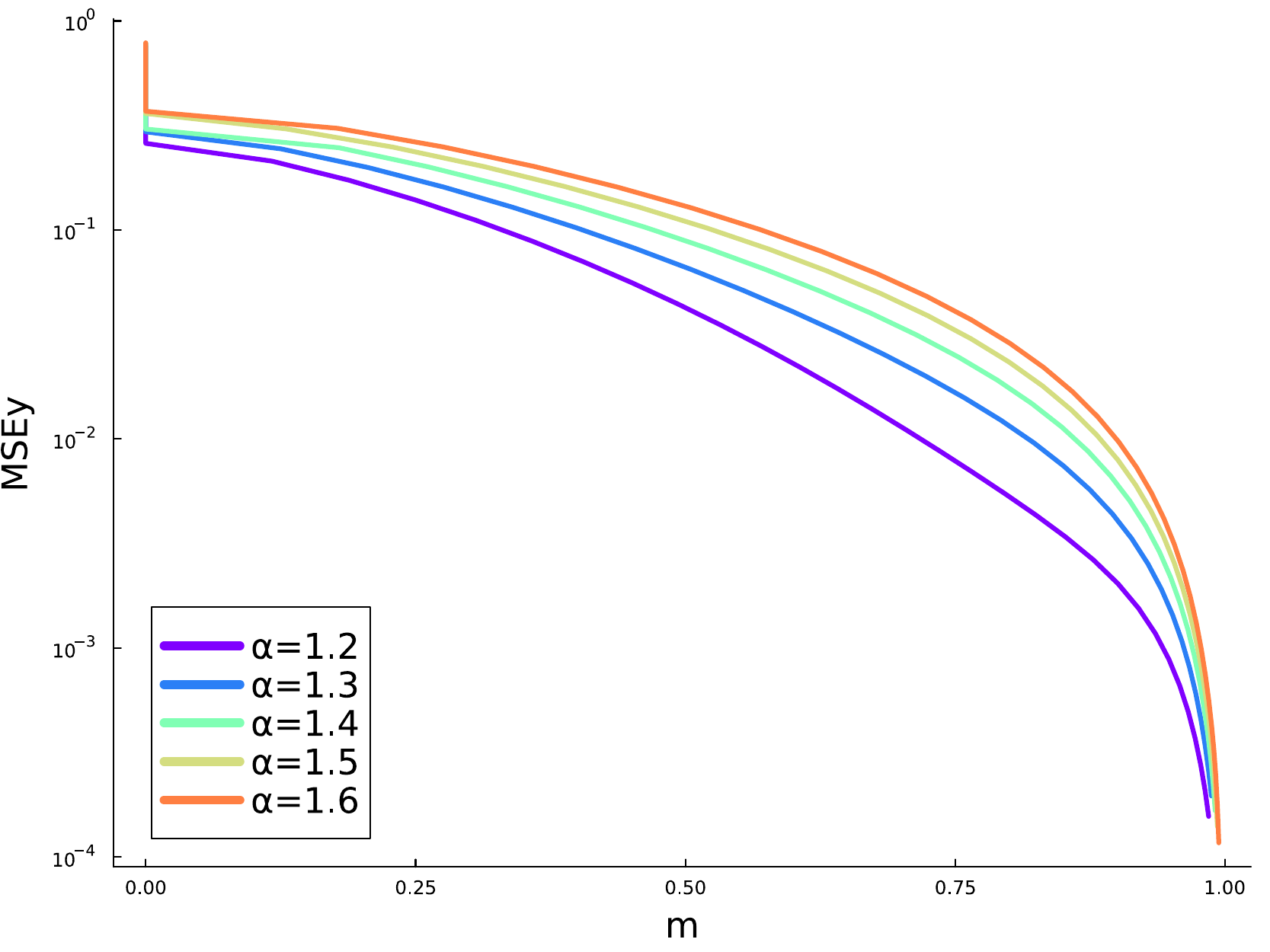}
    \includegraphics[scale=0.3]{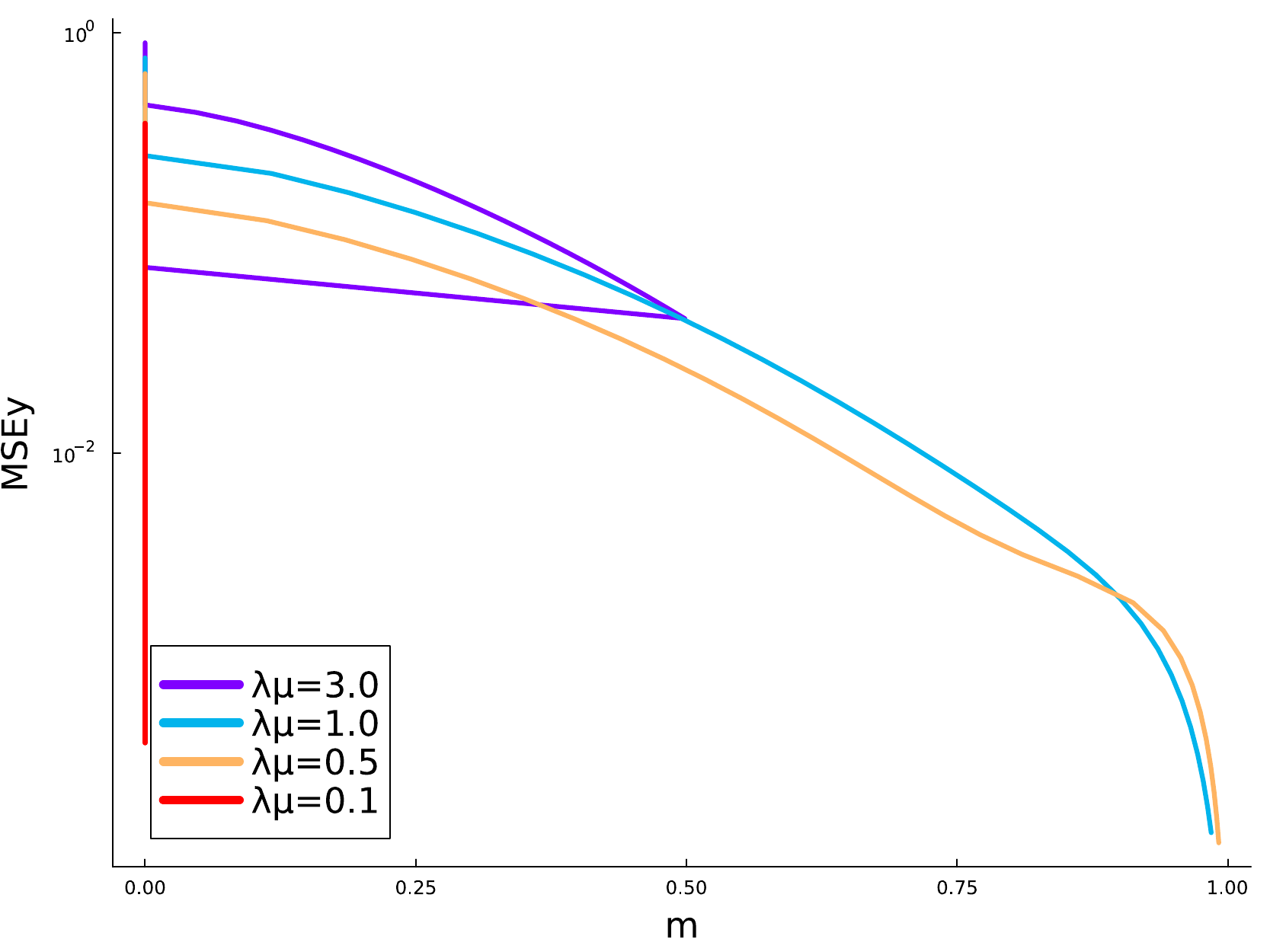}
    \caption{In the \textit{left panel}, we report the mean squared error $MSE_y$ as a function of the magnetization, for the uninformed branch. We fixed the product $\lambda*\mu=1.0$ and we consider different values of of $\alpha\in\{1.2,1.3,1.4,1.5,1.6,1.7\}$. In the \textit{right panel}, we report the same quantities, for a fixed $\alpha=1.2$, varying the product value $\lambda\cdot\mu \in \{0.1,0.5,1.0,3.0\}$.}
    \label{Fig:Annealing}
\end{figure}

The effect of the simultaneous annealing at constant $\mu\lambda$ is presented in left panel of Fig.~\ref{Fig:Annealing}. Compared to Fig.~\ref{Fig:NoLambda} and Fig.~\ref{Fig:Lambda}, an early strong regularization allows the optimization to flow toward the perfect retrieval solution even at very low values of alpha $\alpha \sim \alpha_{BO}$. 
The right panel of Fig.~\ref{Fig:Annealing} shows that it is crucial to keep the product $\mu \cdot \lambda$ small enough to ensure an appropriate value of $\lambda$ for each $\mu$. If $\lambda$ becomes too large at high $\mu$, the bias effect introduced by the regularization becomes evident. 


\section{Numerical Results}
\label{sec:NumericalResults}

The presented replica analysis suggests that partitioning the Phase retrieval problem into two nested optimization levels can benefit optimization. In practice, the Phase selection formulation is also convenient since the internal problem, the optimization of $\mathbf{x}$ given a choice for $\mathbf{S}$, is convex and the solution can be obtained in closed form Eq.~\eqref{eq:OptimizationPhaseSelection}. The external level, the sign selection problem, is instead combinatorial and can be tackled with standard heuristic solvers, such as Simulated annealing \cite{kirkpatrick1983optimization}, Approximate message passing \cite{mezard2009information}. In this study, we also consider Langevin dynamics on the continuous relaxation of the signs, defined below. 

Let us first analyze the behavior of the solvers when the regularization parameter is kept constant. In Sec.\ref{subsec:lambda}, we observed that, for example, at $\alpha = 1.3$ there are ranges of $m$ where both free energy branches are subdominant, making it impossible for a randomly initialized configuration to flow to the signal. Indeed, in this regime, Simulated Annealing (SA) fails to reconstruct $\mathbf{x_0}$. However, the recovery threshold of the algorithm is found to be located at values of $\lambda$ and $\alpha$ higher than the ones predicted by the merging of informed and uninformed branches in the RS analysis, likely due to the onset of Replica Symmetry Breaking. In Appendix \ref{app:RS_stability}, we study the local stability of the RS ansatz, and report the result of the replica analysis under a 1RSB assumption. Exploring the full large deviation analysis in this more complex case is however numerically challenging, and is left for future work.


In the following sections, we explore the retrieval performance of the three algorithms when the simultaneous annealing procedure of $\mu$ and $\lambda$, described in the previous section, is implemented. The pseudo-code for the problem instance generator is reported in App.\ref{app:ProblemGenerator}.

\subsection{Simulated Annealing}
\label{sec:SA}

Simulated Annealing (SA) is a heuristic solver that leverages the Markov Chain Monte Carlo (MCMC) method and a temperature annealing process to control the exploration of the configuration space \cite{kirkpatrick1983optimization}. Note that the application of SA to Phase selection is closely inspired by \citep{obuchi2018statistical}, where a different variable selection problem was considered. The pseudo-code for the SA-based solver is outlined in Algorithm \ref{alg:SA}.
We consider a multiplicative annealing of the inverse temperature:
\begin{align*}
    \mu_t=\mu_{t-1}(1+\epsilon),
\end{align*}
where $\epsilon$ is a hyper-parameter controlling the annealing rate. At each temperature, we perform $T_{A}$ Monte Carlo steps, and then proceed in $\mu$ annealing. It is theoretically established that a ``slow enough'' cooling schedule should allow SA to focus on the global minima of the loss, thereby leading to signal recovery in the retrieval phase of our model.
Simultaneously, we consider an exponential decay of the regularization parameter $\lambda$, from an initial value $\lambda_0$, keeping the product $\mu \lambda$ constant. 

Note that, performing the annealing only after $T_{A}$ steps of MCMC is convenient since it reduces the computational cost of the evaluation of the closed-form optimizer $\boldsymbol{\hat{x}}$ at each iteration. One can rewrite:
\begin{align}
    \boldsymbol{\hat{x}} = \left(A A^{T} + \lambda I_N\right)^{-1}  A^T (\mathbf{S} \odot \mathbf{y}_{t}) = M_{ols} (\mathbf{S} \odot \mathbf{y}_{t}),
    \label{eq:RidgeRegression1}
\end{align}
where $M_{ols}$ needs an update --with a $\mathcal{O}(M^3)$ cost because of the inversion-- only when the regularization intensity changes. Note that the matrix multiplication in the right term of Eq.~\eqref{eq:RidgeRegression1} has a $\mathcal{O}(M^2)$ cost.
Moreover, the impact on the loss of a change of the $\nu$-th sign can be evaluated efficiently after rewriting:
\begin{align}
    MSE_y = \frac{1}{M} \lVert (\mathbb{I} - A^T M_{ols}) (\mathbf{S} \odot \mathbf{y}_{t})\rVert^2_2 = \frac{1}{M} \lVert \mathbf{\Delta_y} \rVert^2_2,
    \label{eq:RidgeRegression2}
\end{align}
and by considering the shift to the vector of errors $\mathbf{\Delta_y}$ induced by the sign flip. The matrix $(\mathbb{I} - A^T M_{ols})$ can be computed only when the regularization value is updated.

\begin{algorithm}[H]
    \caption{Simulating Annealing based solver}
    \label{alg:SA}
    \scriptsize
    \begin{algorithmic}[H]
        \State \textbf{Input}: Phase selection instance ($A$, $\mathbf{y}$)\\ 
        
        \State\textbf{Parameters:} $\epsilon$,  $\mu_0$, $\lambda_0$, $T_{A}$,  \\
    
    \State \textbf{Initialization:}
    \begin{itemize}
        \item The phase vector is uniformly sampled, i.e. $\mathbf{S} \sim \text{Uniform}\left(\{-1,1\}, N\right)$
        \item The inverse temperature is initialized to $\mu_0$
        \item The regularization is initialized to $\lambda_0$
    \end{itemize}\\



    \State{\textbf{Steps of the Algorithm:}}\\
    \noindent \While {$t \leq T$}{
    \begin{itemize}
        \item Choose an index $m$ uniformly $\in \{1,\dots, M\}$
        \item Find the associated shift in the errors: $\mathbf{\delta} = -2S_my_m  (A_{ols}-I_M)_m$
        \item Compute the Energy variation: $\Delta_E = \Delta_E( \mathbf{\Delta_y},  \mathbf{\delta}) = \lVert \mathbf{\Delta_y} + \mathbf{\delta}\rVert_2 - \lVert \mathbf{\Delta_y}\rVert_2$
        \item Accept the move with probability $e^{-\mu \Delta_E}$
        \item Every $T_{A}$ scale:
        \begin{align*}
            \mu = \mu (1 + \epsilon) \qquad \lambda = \frac{\lambda_0 \mu_0}{\mu}\\
        \end{align*}
        and compute
        \begin{align*}
        M_{ols} = \left(A A^{T} + \lambda I_N\right)^{-1} ~ A \qquad A_{ols} = A^{T} O_{ols} \qquad \mathbf{\Delta_y} = (A_{ols}-I_M) (\mathbf{S} \odot \mathbf{y})
        \end{align*}
        
    \end{itemize}
    }
        
    \State \textbf{Return:} $M_{ols} (\mathbf{S} \odot \mathbf{y})$
    \end{algorithmic}
\end{algorithm}



\begin{figure}[H]
    \centering
    \includegraphics[scale=0.4]{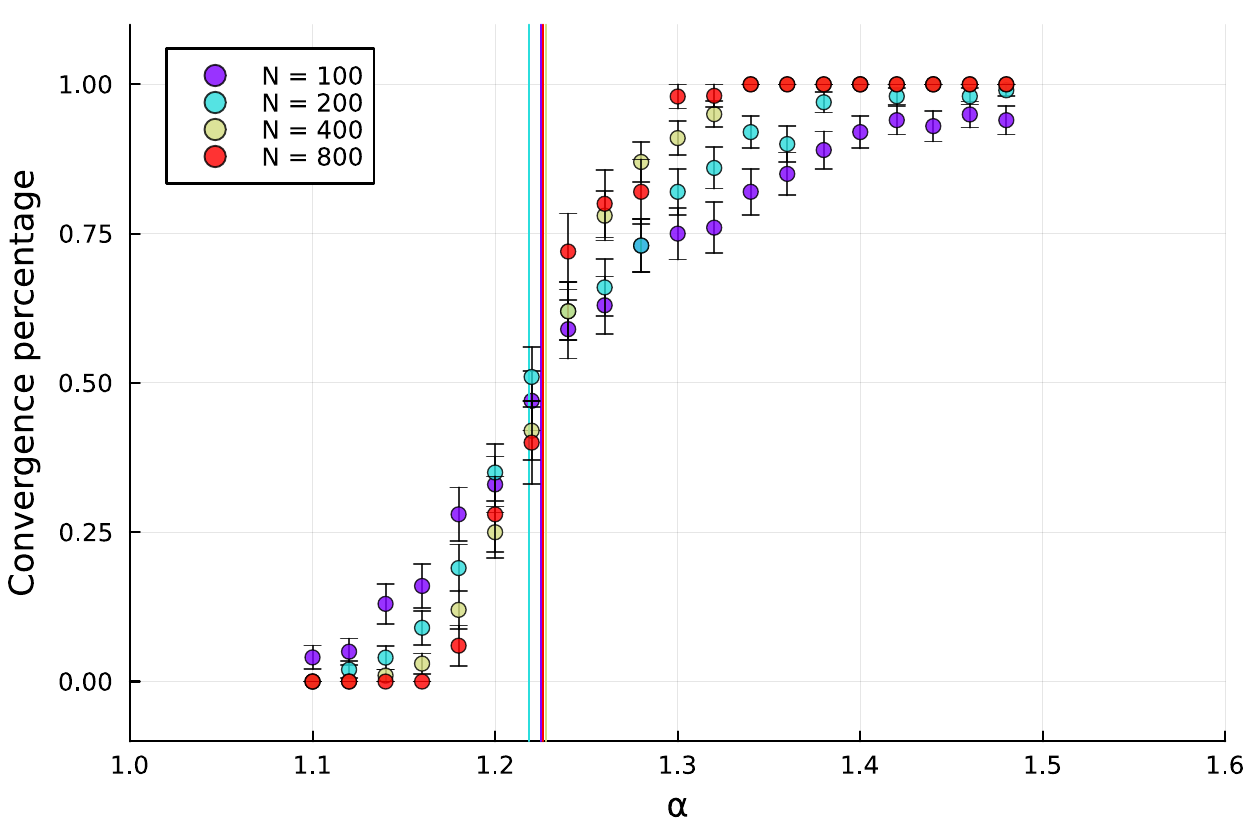}
    \includegraphics[scale=0.4]{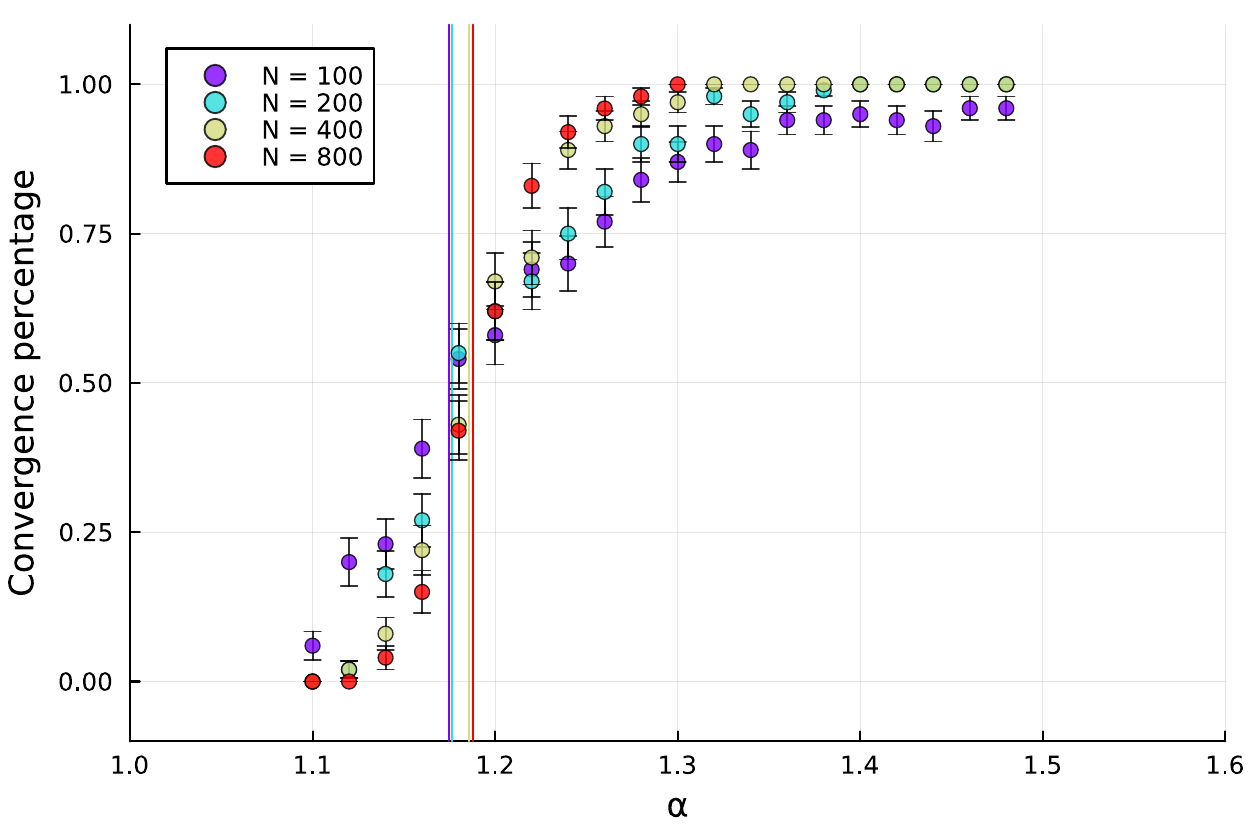}
    \caption{The SA-based solver is employed for solving 100 Phase Selection instances, for $T=10^6, T_A=100, \mu_0=1.0, \lambda_0=1.0$. After $T$ steps the overlap $m$, between $x_0$ and $\hat{x}$, is computed as well as the $MSE_y$, for different values of task complexity, $\alpha\in \text{LinRange}(1.1,1.475,0.025)$, and different values of the sistem size $N\in\{100,200,400,800\}$. We report the percentage of processes able to fully recover $\mathbf{x_0}$, i.e. $m\sim1$ and $MSE_y=0$; and a vertical line representing the $\alpha$ value at which we can solve half of the instances. The annealing schedules are controlled by $\epsilon = 0.1$ (\textit{left}) and $\epsilon = 0.01$ (\textit{right}).}
    \label{Fig:SizeDependenceSA}
\end{figure}

In Fig.\ref{Fig:SizeDependenceSA}, we study the success rate of the SA solver for two values of the annealing rate parameter $\epsilon$, and initial values $\mu_0 = 1$ and $\lambda_0 = 1$. 
As expected, the slower annealing induces a better exploration of the configuration space, leading to a higher recovery probability close to the $\alpha_{BO}$ threshold. 
Notice that the algorithmic threshold $\alpha_{\text{SA}}$, here defined as the dataset size $\alpha$ at which the success probability is one-half, seems to approach the Bayes optimal algorithmic threshold $\alpha_{BO}=1.13$ as larger instances and slower annealing schedules are considered.

\subsection{Approximate Message Passing based solver}
\label{sec:AMP}


Under the assumptions of our model setting, the \emph{a posteriori} mean and variance of the estimator $\mathbf{x}$, for a given value of $\mu$ and $\lambda$, can be estimated via the Approximate Message Passing \cite{mezard2009information} approach. The exact form of the AMP equations for Phase selection, which can be derived by closing on first and second moments the associated relaxed Belief Propagation equations, is specified by defining the so-called input and output channels of the problem:
\begin{align}
    \phi_{in}(A_0, A_1, B; \lambda) &= \frac{1}{2}\log\left(\frac{A_1+\lambda}{A_1+\lambda-A_0}\right)+\frac{B^2}{2(A_1+\lambda-A_0)},\\
    \phi_{out}(V_0, V_1, \omega; y, \mu) &= \frac{1}{2}\log\left(\frac{1+2V_1}{1+2V_1+2\mu V_0}\right) + \log\left(e^{-\mu \frac{(\omega+y)^2}{1+2V_1+2\mu V_0}}+ e^{-\mu \frac{(\omega-y)^2}{1+2V_1+2\mu V_0}}\right). \label{eq:channels}
\end{align}
which are closely related to the entropic and energetic terms in the replica analysis Eq.~\eqref{eq:entropic} and Eq.~\eqref{eq:energetic} \cite{ASP}. The full algorithm is detailed in Fig.~\ref{alg:AMP}. It is known that the State Evolution equations, offering a deterministic low-dimensional description of the trajectories of AMP, are in one-to-one correspondence with the saddle point iterations employed to extremize the replica symmetric free energy \cite{rangan2011generalized}. 

Note that, given the two-level structure of Phase selection, the AMP equations require the estimation of two second moments, yielding a set of recursive equations reminiscent 
of the Approximate Survey Propagation equations, a 1RSB version of AMP \cite{ASP,Saglietti_2020}. This similarity is directly related to the fact that, in Sec.\ref{AnaliticalInvstigation}, the replica trick was applied both to the external and the internal optimization levels.
It is also interesting to notice that the optimization level associated with the sign selection is not immediately evident from the equations since the configuration $\mathbf{S}$ is traced out in the output channel of Eq.\ref{eq:channels}. At large values of $\mu$, only the largest term inside the logarithm will give a finite contribution, effectively corresponding to polarization on a specific sign choice. Moreover, because of the Gaussian nature of the inputs, in the AMP approach, we are not required to use the closed-form solution of the internal problem, avoiding the costly matrix inversion.  

Here, we aim to turn AMP into a Phase selection solver by introducing our simultaneous annealing procedure on $\mu$ and $\lambda$ during the message-passing iterations, forcing the measure to focus on a single configuration. This procedure should approximate state-following as long as the annealing is not too abrupt. Again, we use a hyper-parameter $\epsilon$ to control the $\mu$-annealing rate. The full heuristic is detailed in Alg.\ref{alg:AMP}.
After convergence, one can use the fixed point estimate $\mathbf{x}^{\star}$ to compute the estimated labels $A \mathbf{x}^\star$ as well as the Phase Selection loss. 

It's important to highlight that, if the training inputs were not i.i.d. Gaussian distributed, as is often the case in practical inference tasks, the development of correlations could undermine the reliability of the Approximate Message Passing (AMP) algorithm, potentially leading to issues with convergence or inaccurate loss estimates. In the case of a generalized covariance, the vector-AMP approach \cite{rangan2018vectorapproximatemessagepassing} should correct for the inaccurate assumptions of our implementation, but its derivation and the associated analysis are beyond the scope of this work.

\begin{figure}[htbp]
    \centering
    \includegraphics[scale=0.4]{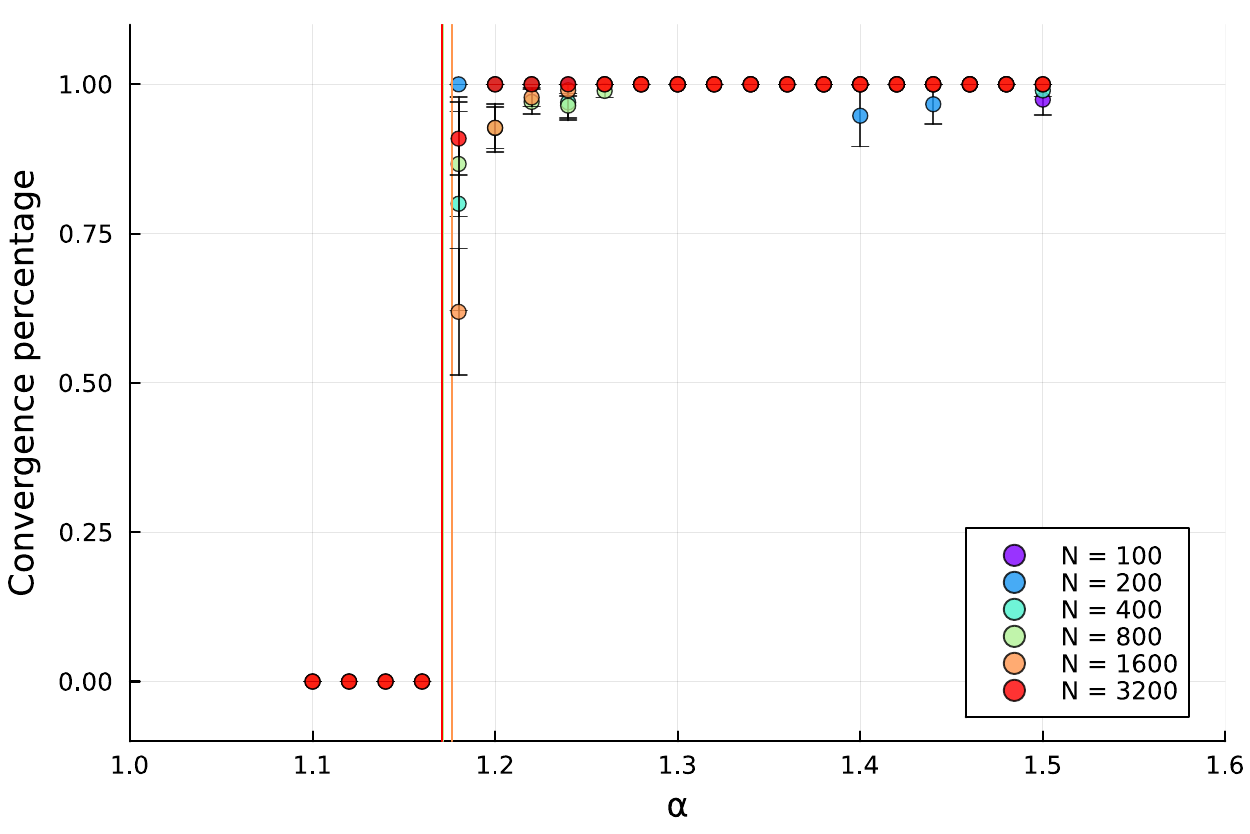}
    \includegraphics[scale=0.4]{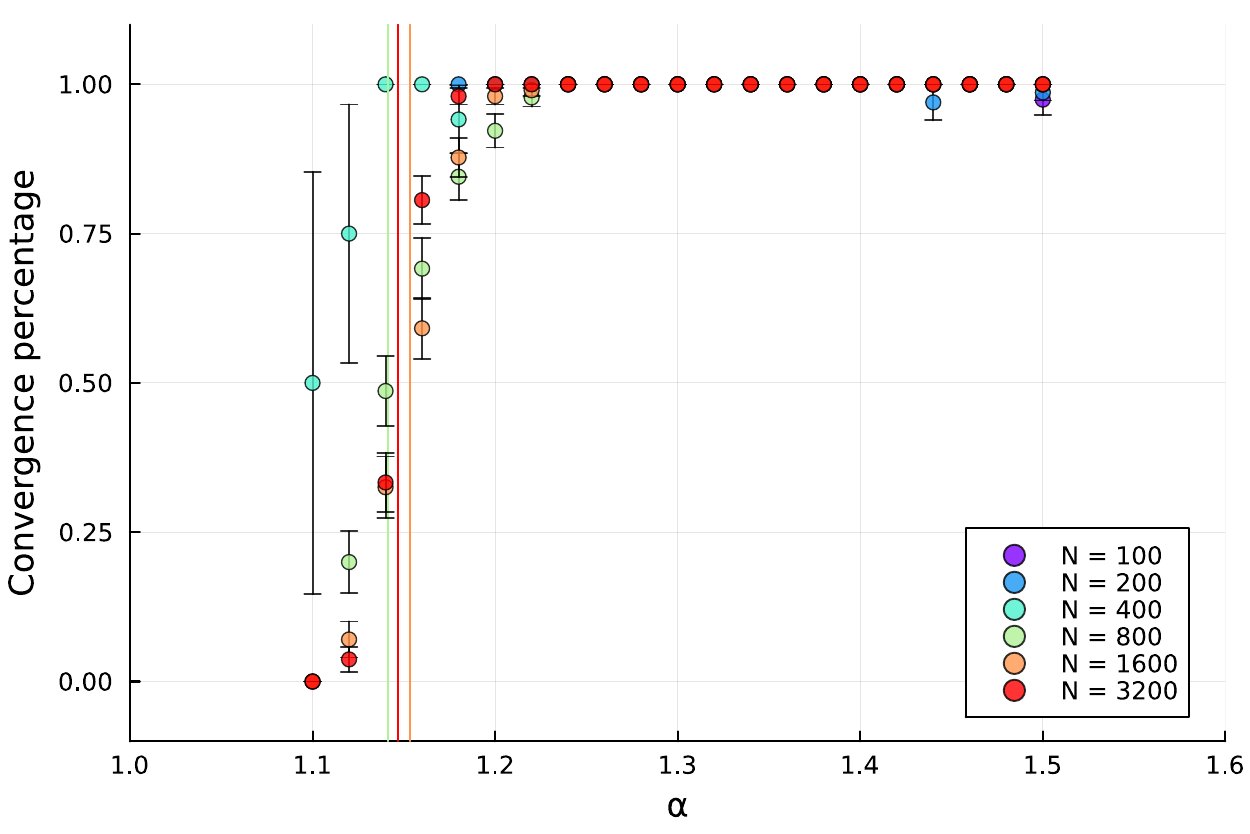}
    \caption{The AMP-based solver is employed for on $100$ Phase Selection instances, for $T=2\cdot 10^5, \mu_0=1000.0, \lambda_0=1.0$. After $T$ steps the overlap $m$, between $x_0$ and $\hat{x}$, is computed as well as the $MSE_y$, for different values of task complexity, $\alpha\in \text{LinRange}(1.1,1.475,0.025)$, and different values of the sistem size $N\in\{100,200,400,800, 1600, 3200\}$. We report the percentage of runs able to fully recover $\mathbf{x_0}$, i.e. $m\sim1$ and $MSE_y=0$; and a vertical line representing the $\alpha$ value at which we can solve half of the instances. The annealing rate is controlled respectively by $\epsilon = 0.1$ (\textit{left}) and $\epsilon = 0.01$  (\textit{right}).}
    \label{Fig:SizeDependenceASP}
\end{figure}
.
\vspace{-1.5cm}
\begin{algorithm}[H]
\caption{\\ Approximate Message Passing based solver}
\label{alg:AMP}
  \scriptsize
\begin{algorithmic}[t]
    \State \textbf{Input:} Phase Selection instance($A$, $\mathbf{y}$)\\

    \State \textbf{Parameters:} $T, \mu_0, \lambda_0, \alpha, \epsilon, \psi, \eta$\\
    
    \State \textbf{Initialize:} 
    \begin{itemize}
        \item Construct the AMP-Graph associated to Phase Selection instance, i.e. $\mathbf{g}_{out}$, $A_0$, $A_1$, $\mathbf{B}_1$, $\mathbf{\Delta}_{0}$, $\mathbf{\Delta}_{1}$, $\boldsymbol{V_{0}}$, $\boldsymbol{V_{1}}$, $\boldsymbol{\omega}$
        \item The regularization and the inverse temperature are initialized to $\lambda_0$ and $\mu_0$
        \item The annealing parameter are so defined $\mu_{A} = 1+\epsilon$.
    \end{itemize}\\
    
    \State \textbf{Steps of the Algorithm}\\

    \noindent\While{$t<T$ or $\Delta_t<\eta$}{
        \begin{align*}
        V_{0,\nu}^t &= \sum_i  \left(A_i^{\nu}\right)^2 \Delta_{0,i}^{t-1} (1-\psi) + V_{0,\nu}^{t-1} \psi \\
        V_{1,\nu}^t &= \sum_i  \left(A_i^{\nu}\right)^2 \Delta_{1,i}^{t-1} (1-\psi) + V_{1,\nu}^{t-1} \psi\\
       \mathbf{\omega}_{\nu}^t &= \sum_i A_i^{\mu}x_i^{t-1} - \left(\sum_i  \left(A_i^{\nu}\right)^2 \Delta_{0,i}^{t-1} + \frac{\sum_i  \left(A_i^{\nu}\right)^2 \Delta_{1,i}^{t-1}}{\mu}\right) g_{out,\nu}^{t-1}(1-\psi) + \mathbf{\omega}_{\nu}^{t-1}\psi   
        \end{align*}


    \begin{align*}
    g_{out,\nu}^t &= \partial_{\omega^t_{\nu}}\phi_{out}(\omega^t_{\nu}, y_{\nu}, V_{0,\nu}^t,V_{1,\nu}^t, \mu)\\
    A_{0}^{t} &=  2\mu \sum_\nu \left(A_i^{\nu}\right)^2 \partial_{V^t_{1,\nu}}\phi_{out}(\mathbf{\omega}^t_{\nu}, y_{\mu}, V_{0,\mu}^t,V_{1,\nu}^t, \mu) - \left(g^{t-1}_{out, \nu}\right)^2\\
    A_{1}^{t} &=  \sum_\nu  \left(A_i^{\nu}\right)^2\partial^2_{\omega^t_{\nu}}\phi_{out}(\omega^t_{\nu}, y_{\nu}, V_{0,\nu}^t,V_{1,\nu}^t, \mu) + 2\mu\partial_{V^t_{1,\nu}}\phi_{out}(\mathbf{\omega}^t_{\nu}, y_{\nu}, V_{0,\mu}^t,V_{1,\nu}^t, \mu) - \left(g^{t-1}_{out, \nu}\right)^2 \\
    B_i^{t} &= \sum A_{i}^{\nu} g^t_{out, \nu} - x_i^{t-1}\left(A_{0}^{t}-A_{1}^{t}\right)
    \end{align*}

    \begin{align}
    x_i^{t} &= \partial_{B_i^t} \phi_{in} \biggl(B_i^t, A_0^t, A_1^t,\lambda \biggr)\\
    \Delta_{0,i}^t &= -2\partial_{A^t_1} \phi_{in}\biggl(B_i^t, A_0^t, A_1^t,\lambda \biggr) - \left(x_i^{t}\right)^2\\
    \Delta_{1,i}^t &= \partial^2_{B_i^t} \phi_{in}\biggl(B_i^t, A_0^t, A_1^t,\lambda \biggr) - \Delta_{0,i}^t
    \end{align}
    
    
    \begin{align*}
         \mu \cdot = \mu_{A} \qquad \lambda = \lambda_0 \mu_0/\mu
    \end{align*}
    }
    
    \State \textbf{Return:} $A$, $\Delta$
\end{algorithmic}
\end{algorithm}

Fig.\ref{Fig:SizeDependenceASP} displays the recovery probability for two values of the annealing rate parameter $\epsilon$. Also in this case, the heuristic seems to approach the $\alpha_{BO}$ threshold in the limit of vanishing annealing rate. 

\subsection{Langevin dynamics based solver}

Finally, we explore a more heuristic approach, relaxing the combinatorial problem of the sign selection into a continuous optimization problem. We replace the binary variables $\mathbf{S}$ with a continuous vector $\mathbf{\theta}\in\mathbb{R}^M$. The continuous optimization of $\mathbf{\theta}$ an be performed via Langevin dynamics over the loss:
\begin{align*}
    L(\mathbf{\theta}, A_{\text{ols}}, \mathbf{y}, \mathbf{\lambda}, N) = \lVert \mathbf{\hat{y}} - A_{\text{ols}} \cdot \mathbf{\hat{y}} \rVert^2_2,
\end{align*}
where $\mathbf{\hat{y}} = \mathbf{\theta} \odot |\mathbf{y}|$ and $A_{\text{ols}} = A^T M_{ols}$, with $M_{ols}$ defined as in Eq.~\eqref{eq:RidgeRegression1}. Note that, during the optimization of the continuous signs, it is useful to bound their values inside the hypercube, by rescaling the gradient (Eq.~\eqref{eq:rescale_grad}) and clipping the values of $\mathbf{\theta}$, $\theta_i\in[-1+1/N, 1-1/N]$.
Similarly to the SA-based solver, we consider a multiplicative annealing in the inverse temperature $\mu$, affecting the magnitude of the Langevin Dynamics noise. The product $\lambda \cdot \mu $ is kept constant and updated once every $T_{A}$, together with $M_{ols}$.  The pseudo-code of the Langevin dynamics-based solver is reported in Alg.\ref{alg:GD}.

\begin{algorithm}[H]
    \caption{Langevin dynamics based solver}
    \label{alg:GD}
    \scriptsize
    \begin{algorithmic}[H]
        \State \textbf{Input}: Phase selection instance ($A$, $\mathbf{y}$)\\ 
        
        \State\textbf{Parameters:} $\epsilon$,  $\beta_0$, T, $T_{A}$, $\lambda_0$, $\eta$ \\
    
    \State \textbf{Initialization:}
    \begin{itemize}
        \item The continuous phase vector is uniformly sampled, i.e. $\boldsymbol{\theta} \sim \text{Uniform}\left(-1,1\right)^N$
        \item The inverse temperature is initialized to $\mu_0$
        \item The regularization is initialized to $\lambda_0$
        \item The annealing parameter are so defined $\mu_{A} = 1+\epsilon$.
    \end{itemize}\\

    \State{\textbf{Steps of the Algorithm:}}\\
    \noindent \While {$t \leq T$}{
    \begin{itemize}
        \item Compute and randomize gradient loss 
        \begin{align}
        g(\mathbf{\theta}, A_{\text{ols}}, \mathbf{y}, \mathbf{\lambda}, N) = \nabla_{\boldsymbol{\theta}} L(\mathbf{\theta}, A_{\text{ols}}, \mathbf{y}, \mathbf{\lambda}, N) + \sqrt{\frac{2}{\mu\eta}}\mathbf{z}    \qquad
        \mathbf{z}\sim \mathcal{N}_N(0,1)
        \end{align}.
        \item Normalize the gradient 
        \begin{align}\label{eq:rescale_grad}
            g(\mathbf{\theta}, A_{\text{ols}}, \mathbf{y}, \mathbf{\lambda}, N) = g(\mathbf{\theta}, A_{\text{ols}}, \mathbf{y}, \mathbf{\lambda}, N) \odot  (\mathbf{1} - \boldsymbol{\theta}^2) 
        \end{align}
        \item Update $\boldsymbol{\theta}$ and check $\boldsymbol{\theta} \in [-1+1/N, 1-1/N]$
        
        \item Every $T_{A}$ scale:
        \begin{align*}
        \mu \cdot= \mu_{a}\qquad \lambda = \lambda_0\mu_0/\mu\\
        \end{align*}
        and compute
        \begin{align*}
        O_{ols} = \left(A A^{T} + \lambda I_N)\right)^{-1} ~ A \qquad A_{ols} = A^{T} O_{ols} \qquad \mathbf{\Delta_S} = (A_{ols}-I_M) (\mathbf{S} \odot \mathbf{y})
        \end{align*}
        
    \end{itemize}
    }
        
    \State \textbf{Return:} $O_{ols} * (\mathbf{s} \cdot \mathbf{y})$
    \end{algorithmic}
\end{algorithm}

\begin{figure}[h]
    \centering
    \includegraphics[scale=0.4]{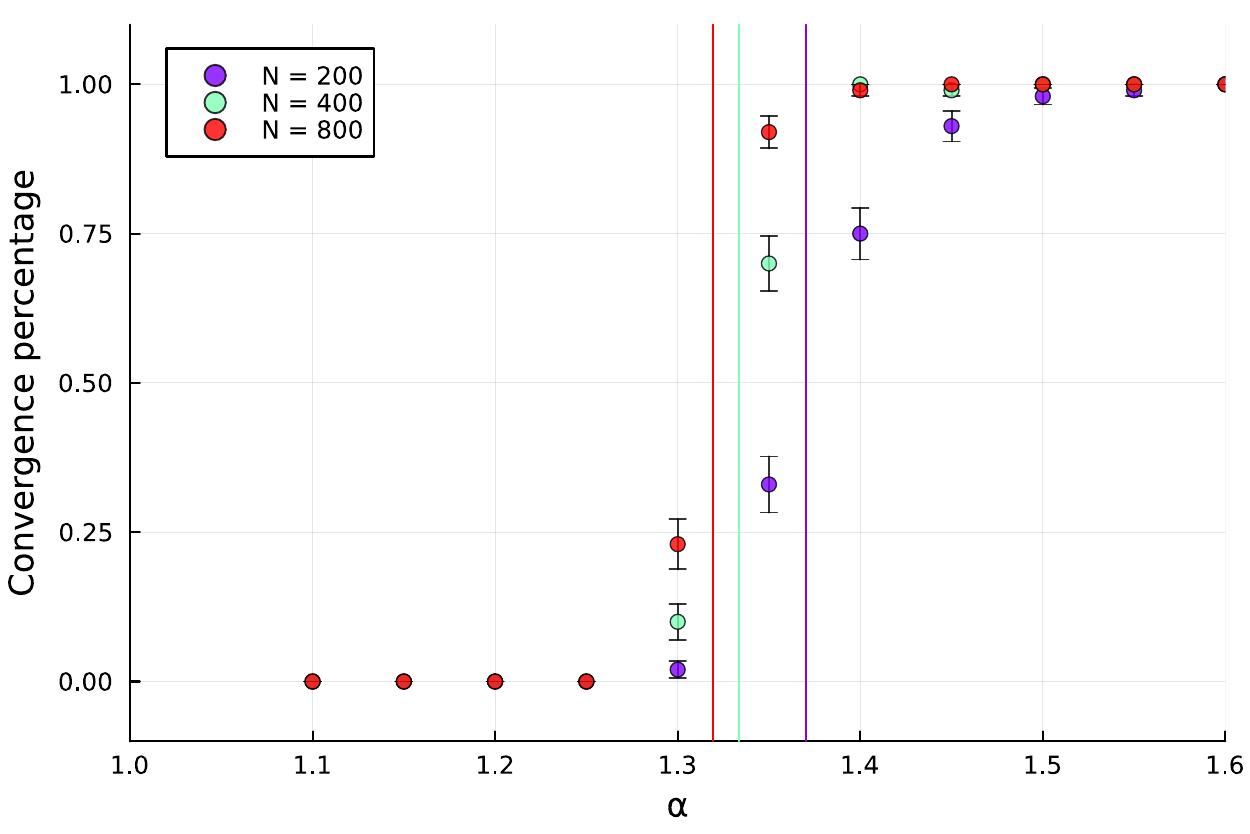}
    \includegraphics[scale=0.4]{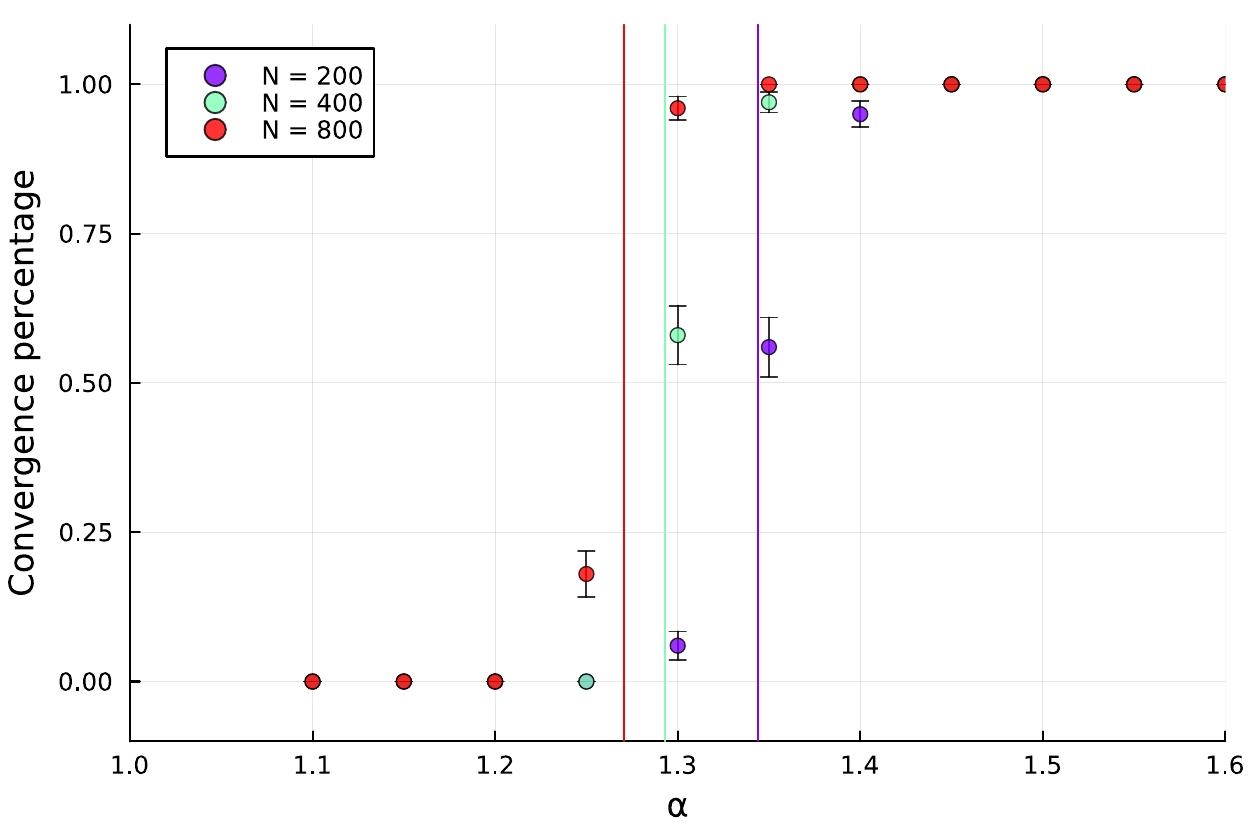}
    \caption{The GD-based solver is employed for solving 100 Phase Selection instances, for $T=10^6, T_A=100, \mu_0=1.0, \lambda_0=0.75$. After $T$ steps the overlap $m$, between $x_0$ and $\hat{x}$, is computed as well as the $MSE_y$, for different values of task complexity, $\alpha\in \text{LinRange}(1.1,1.475,0.025)$, and different values of the sistem size $N\in\{200,400,800\}$. We report the percentage of processes able to fully recover $\mathbf{x_0}$, i.e. $m\sim1$ and $MSE_y=0$; the vertical lines represent the $\alpha$ values at which half of the runs achieve recovery. The annealing schedules are $\epsilon = 0.1$ (\textit{left}) and $\epsilon = 0.01$ (\textit{right}).}
    \label{Fig:SizeDependenceSGD}
\end{figure}

Fig.~\ref{Fig:SizeDependenceSGD} shows the empirical probability of solving a phase selection instance with the Langevin dynamics approach. Once again, the effect of the annealing rate \( \epsilon \) is evident: slower schedules lead to better results. 
By comparing Fig. \ref{Fig:SizeDependenceSGD} with Fig. \ref{Fig:SizeDependenceSA} and Fig.\ref{Fig:SizeDependenceASP}, it is clear that the retrieval performance of the Langevin dynamics-based scheme is the poorest among the considered solvers, due to the heuristic nature of the continuous relaxation strategy probably limits the performance. However, note that this approach is still able to achieve recovery with a gradient-based approach in regimes where typically gradient descent remains stuck in local minima orthogonal to the signal. 

\section{Discussion} \label{sec:discussion}

In this work, we have introduced and analyzed the \emph{Phase Selection} formulation as a means to disentangle the combinatorial and convex continuous components of the Phase Retrieval problem. By separating the hard combinatorial subproblem—retrieving the missing measurement signs—from the convex regression task, we have demonstrated that heuristic optimization methods can solve harder instances, approaching the Bayes-optimal (BO) threshold. Moreover,
our findings provide a refined explanation for the effectiveness of regularization in phase retrieval, an observation previously noted in the literature \cite{ma2018approximate, Saglietti_2020}, but here linked more explicitly to the free energy landscape and the bias toward fitting measurements with larger magnitudes. Our statistical physics analysis, grounded in large deviation theory and replica methods, also indicates that the introduction of an adaptive regularization scheme can accelerate convergence by biasing optimization towards informative solutions at early stages while avoiding systematic errors at later stages. 

An open direction for future research is to investigate whether this type of decomposition approach could be beneficial in other high-dimensional optimization problems, particularly those where a discrete structure (e.g., spin-glass-like variables) is intertwined with continuous optimization. Potential areas of interest include inference in graphical models with hidden variables, clustering in kernelized spaces, and even aspects of reinforcement learning where discrete action selection is coupled with continuous value optimization. For example, in learning problems involving weight symmetries—such as neural networks—breaking permutation symmetry explicitly could lead to improved training dynamics by reducing the effective search space. This is reminiscent of how certain architectural choices, such as weight tying in convolutional networks or specialized initialization schemes, facilitate optimization in deep learning models \cite{EscapingMediocrity,Stability_Urbani}. More generally, the interplay between regularization and combinatorial structure in optimization warrants further theoretical and empirical investigation.



\bibliographystyle{unsrt}
\bibliography{biblio.bib}

\newpage

\appendix

\section{Analytical results}

\subsection{Stabilities analysis}\label{app:stabilities}

In this section, we provide the detailed computation of the distribution of stabilities presented in Sec.\ref{subsec:Stabilities}. Stabilities represent the measurement amplitudes given by a signal estimate $\mathbf{x}$:

\begin{align*}
    \frac{\mathbf{x}\cdot\boldsymbol{A}^{\nu}}{\sqrt{N}}. 
\end{align*}
The distribution of stabilities is defined as the expected value of the Dirac delta $\delta(\cdot)$ of the difference between the magnitude of the measurement $\Delta$ and the stability:  

\begin{align*}
    P(\Delta) = \left\langle \delta\left(\Delta - \frac{\mathbf{x}\cdot\boldsymbol{A}^{1}}{\sqrt{N}}\right) \right\rangle=\mathbb{E}_{A,\mathbf{x_0}}\left[\frac{\sum_{\left\{ S\right\} }e^{\phi\sum_{\nu=1}^M S_{\nu}-\mu \mathcal{H} \left(\mathbf{S}\,|\,A,\mathbf{x_0}\right)} \delta\left(\Delta - \frac{\mathbf{x}\cdot\boldsymbol{A}^{1}}{\sqrt{N}}\right) }{\sum_{\left\{ S\right\} }e^{\phi\sum_{\nu=1}^M S_{\nu}-\mu \mathcal{H} \left(\mathbf{S}\,|\,A,\mathbf{x_0}\right)}}\right],
    \end{align*}

The presence of the patterns $\mathbf{A}^{\nu}$ at the denominator makes this computation hard, however, we can introduce two sets of replicas $a\in\{1,\dots,n\}$ and $c\in\{1,\dots,\beta\}$, with $\beta=\mu/\rho$. By replicating the procedure reported in Sec.\ref{sec:large deviation}, we can rewrite the distribution of stabilities as:

\begin{align*}
    P(\Delta) &= \lim_{n \to 0}	\mathbb{E}_{A,\mathbf{x_0}}\sum_{\left\{ S^{a}\right\} }\prod_{a}e^{\phi\sum_{\nu}S_{\nu}^{a}}\times\\
    &\times\left[\int\prod_{a=1}^{s}\prod_{c=1}^{\beta}d\mathbf{x}_{ac} P\left(\mathbf{x}_{ac}\right)\prod_{\nu=1}^{\alpha N}\exp\left(-\frac{\mu}{\beta}\sum_{ac}\left(S_{\nu}^{a}\frac{A^{\nu}\cdot \mathbf{x_0}}{\sqrt{N}}-\frac{A^{\nu}\cdot \mathbf{x}_{ac}}{\sqrt{N}}\right)^{2}-\sum_{ac}\frac{\lambda\mu}{2\beta}\lVert \mathbf{x_{ac}} \lVert_2^2.\right)\delta\left(\Delta - \frac{A^{1}\cdot \mathbf{x}_{1,1}}{\sqrt{N}}\right)\right],
\end{align*}
By construction, it is clear that the replica $a=1$ and $c=1$, and the pattern index $\nu=1$ are special and must be considered separately in the computation. For this reason, we need to separate their contribution from the others:

\begin{align*}
    P(\Delta) &= \lim_{n \to 0}	\mathbb{E}_{A,\mathbf{x_0}} \int\prod_{a=1}^{s}\prod_{c=1}^{\beta}d\mathbf{x}_{ac}P\left(\mathbf{x}_{ac}\right) \delta\left(\Delta - \frac{A^{1}\cdot \mathbf{x}_{1,1 }}{\sqrt{N}}\right) \times \\
    &\times\sum_{S^{1}}e^{\phi S_{1}^{1}} \exp\left(-\frac{\mu}{\beta}\left(S^{1}\frac{A^{1}\cdot \mathbf{x_0}}{\sqrt{N}}-\Delta\right)^{2}-\frac{\lambda\mu}{2\beta}\lVert \mathbf{x_{1,1}} \lVert_2^2.\right) \times\\
    &\times \prod_{c>1} \exp\left(-\frac{\mu}{\beta}\sum_{c>1}\left(S_{1}^{1}\frac{A^{1}\cdot \mathbf{x_0}}{\sqrt{N}}-\frac{A^{1}\cdot \mathbf{x}_{1c}}{\sqrt{N}}\right)^{2}-\sum_{c>1}\frac{\lambda\mu}{2\beta}\lVert \mathbf{x_{1c}} \lVert_2^2.\right)\times\\
    &\times \sum_{\left\{ S^{a}\right\} }\prod_{a>1,c}e^{\phi S_{1}^{a}} \exp\left(-\frac{\mu}{\beta}\sum_{a>1,c}\left(S_{1}^{a}\frac{A^{1}\cdot \mathbf{x_0}}{\sqrt{N}}-\frac{A^{1}\cdot \mathbf{x}_{ac}}{\sqrt{N}}\right)^{2}-\sum_{a>1,c}\frac{\lambda\mu}{2\beta}\lVert \mathbf{x_{ac}} \lVert_2^2.\right)\times\\
    &\times \sum_{\left\{ S^{a}\right\} }\prod_{a,c}e^{\phi\sum_{\nu}S_{\nu}^{a}} \prod_{\nu=2}^{\alpha N} \exp\left(-\frac{\mu}{\beta}\sum_{ac}\left(S_{\nu}^{a}\frac{A^{\nu}\cdot \mathbf{x_0}}{\sqrt{N}}-\frac{A^{\nu}\cdot \mathbf{x}_{ac}}{\sqrt{N}}\right)^{2}-\sum_{ac}\frac{\lambda\mu}{2\beta}\lVert \mathbf{x_{ac}} \lVert_2^2.\right).
\end{align*}
where we choose a flat prior for $\mathbf{x}_{ac}$, a normal prior for the hidden signal and we assume the patterns to be distributed normally. Using the Replica Symmetric ansatz reported in Sec.\ref{sec:large deviation} and averaging over the patterns, one can rewrite the expression for the distribution of stabilities as\cite{Engels}: 

    \begin{align*}    P(\Delta)&=n\int\mathcal{D}z_{0}\int\mathcal{D}u_{0}\frac{\sum_{s=\{\pm1\}}e^{-s\phi}\int Dz_{1}\left(\int\frac{du}{\sqrt{2\pi}}e^{-\rho\left(\frac{u^{2}}{2}+\omega(\delta q, s, u, u_0, q_0,q_1, m, z_0, z_1)^{2}\right)}\right)^{\mu/\rho-1}}{\sum_s e^{-s\phi}\int Dz_{1}\left(\int\frac{du}{\sqrt{2\pi}}e^{-\rho\left(\frac{u^{2}}{2}+\omega(\delta q, s, u, u_0, q_0,q_1, m, z_0, z_1)^{2}\right)}\right)^{\mu/\rho}}\times\\
    &\times \int\frac{du}{\sqrt{2\pi}}e^{-\rho\left(\frac{u^{2}}{2}+\omega(\delta q, s, u, u_0, q_0,q_1, m, z_0, z_1)^{2}\right)}\delta\left(\Delta- \left(u\sqrt{\delta q}+z_1\sqrt{q_1-q_0}+\sqrt{q_0-m^2}z_0+m u_0\right)\right)
    \end{align*}
    where
    \begin{align*}
    \omega(\delta q, \tilde{q}, s, u, u_0, q_0,q_1, m, z_0, z_1)&= u\sqrt{\delta q}+s u_0 \sqrt{\tilde{q}}+\sqrt{q_{0}-m^{2}}z_{0}+mu_{0}+\sqrt{q_{1}-q_{0}}z_{1}
\end{align*}

Then by taking the limit for vanishing number of replicas $n\to 0$ $\beta=\mu/\rho\to 0$, and manipulating, it is possible to write the expression for the distribution of stabilities as

\begin{align*}
    P(\Delta) = \int \mathcal{D}z_0 \int \mathcal{D}u_0 \frac{\frac{(1+2\delta q)}{\sqrt    2\pi(q_1-q_0)}\sum_{s\in\{-1,1\}}\frac{e^{-\frac{z^*_+(\Delta, \delta_q, q_1, q_0, m, s)^2}{2}}}{\sqrt{2\pi(q_1-q_0)}}e^{-s\phi}e^{-\mu\left(su_0+\Delta \right)^2(1+2\delta q)}}{\sqrt{\frac{1+2\delta q}{(1+2\delta q) + 2\mu (q_1-q_0)}} e^{-\mu\frac{\left(\sqrt{q_0-m^2}z_0 +m u_0\right) ^2 + u_0^2}{(1+2\delta q) + 2\mu (q_1-q_0)}}2\cosh\left( \phi + 2\mu \frac{\left(\sqrt{q_0-m^2}z_0 +m u_0\right) u_0}{(1+2\delta q) + 2\mu (q_1-q_0)} \right)},
\end{align*}
with 
$$
z_1^{*}(s)=\frac{2\delta qs\sqrt{\tilde{q}}u_{0}-\sqrt{q_{0}-m^{2}}z_{0}-mu_{0}+\left(1+2\delta q\right)\Delta}{\sqrt{q_{1}-q_{0}}}.
$$
The distribution of stabilities, associated with the correct sign choice is obtained by choosing $s=1$, in the previous expression.

\subsection{Stability of the RS solution}\label{app:RS_stability}

As discussed in Sec.\ref{subsec:RSAnsatz}, the Replica Symmetric (RS) ansatz, which assumes uniform behavior across all replicas, may fail to encapsulate the system's complexity fully. To evaluate whether a more refined ansatz incorporating Replica Symmetry Breaking (RSB) is necessary, we analyze the nature of the RS solutions, which may correspond to either local or global optimizers of the free entropy $\Phi$ \cite{PhysRevX.9.011020, Stability_Urbani}. Specifically, if the RS solutions are local optimizers of the free entropy, large enough perturbations to the system lead to modifications in the solutions of the saddle point equations. Conversely, no perturbation alters the RS solution if the RS solutions represent a global optimum. While local stability guarantees that the RS phase is robust against small perturbations, it does not imply that the RS phase constitutes the most stable (i.e., globally optimal) configuration. In the following the stability nature of the solution described in Sec.\ref{subsec:RSAnsatz} is addressed.

\subsubsection{1-step Replica Symmetry Breaking}
\label{subsec:1RSB}

To start the analysis of the RS solution's nature, it is natural first to question whether it constitutes a global optimizer of the free entropy $\Phi$. A rigorous assessment of its global stability requires replacing the Replica Symmetric (RS) ansatz in Eq.\eqref{eq.FreeEnergy} with a 1-Replica Symmetry Breaking (1RSB) ansatz. While the RS ansatz assumes identical interactions among all replicas, the 1RSB ansatz introduces a hierarchical structure, grouping replicas into $N/\tau$ non-overlapping clusters, each containing $\tau$ elements ($\tau$ is called \textit{Parisi parameter}) \cite{Replica,Replica2}. Interactions between replicas within the same group differ from those between replicas in different groups. The new form of the overlaps $q_{ac,bd}$ and $m_{ac}$ follows: 
\begin{gather}
    \begin{aligned}
    q_{ac,bd}&=\begin{cases}
    q_{2} & a=b,c=d\\
    q_{1} & a=b,c>d\\
    q_{0}^{d} & a>b,B_{a}=B_{b}\\
    q_{0}^{0} & a>b,B_{a}\neq B_{b}
    \end{cases};\\
    m_{ac}&=m.
    \end{aligned}  
    \label{eq:1RSBAnsatz}
\end{gather}
In Eq.\eqref{eq:1RSBAnsatz}, $\{B_a\}_{a\in\{1,\dots,N/\tau\}}$ is the set of different groups of replicas. 

The new structure of interacting replicas can be used to compute the 1RSB free entropy expression. The procedure is the same as described in Sec.\ref{sec:large deviation}. In the zero temperature limit, $\rho \to \infty$, we impose the standard rescaling of the order parameters:

\begin{gather}
    \begin{aligned}
        q_{2}&=q_{1}+\frac{\delta q}{\rho};\\
        \hat{q}_{1}-\hat{q}_{2}&=\frac{\rho}{\mu}\delta\hat{q};\\
        \hat{q}_{1/0}&=\left(\frac{\rho}{\mu}\right)^{2}\hat{q}_{1/0};\\
        \hat{m}&=\frac{\rho}{\mu}\hat{m}.
    \end{aligned}
\end{gather}

By plugging the rescaled order parameters in Eq.\eqref{eq:GenFreeEntropy} we can compute the final expression for the 1RSB versions of the Entropic, Energetic, and Interaction terms:

\begin{gather}
    \begin{aligned}
        &g_I= \lim_{n\to0}\frac{G_{i}}{n}	=\frac{1}{2}\left(\delta\hat{q}q_{1}-\frac{\hat{q}_{1}\delta q}{\mu}-\hat{q}_{1}q_{1}-\left(\tau-1\right)\hat{q}_{0}^{d}q_{0}^{d}+\tau\hat{q}_{0}^{0}q_{0}^{0}-2\hat{m}m\right);\\
        &g_S= \lim_{n\to0}\frac{\ln G_{S}}{n}	=-\frac{1}{2}\log\left[\left(1-\frac{\hat{q}_{1}-\hat{q}_{0}^{d}}{\delta\hat{q}}\right)\left(1-\frac{\tau(\hat{q}_{0}^{d}-\hat{q}_{0}^{0})}{\left(\delta\hat{q}-\left(\hat{q}_{1}-\hat{q}_{0}^{d}\right)\right)}\right)\right]+\left[\frac{\left(\hat{q}_{0}^{0}+\hat{m}^{2}\right)}{2\left(\left(\delta\hat{q}-\left(\hat{q}_{1}-\hat{q}_{0}^{d}\right)\right)-\tau\left(\hat{q}_{0}^{d}-\hat{q}_{0}^{0}\right)\right)}\right];\\
        &g_E= \lim_{n\to0}\frac{\ln G_{E}}{n}	=\frac{1}{2}\log\left(\frac{1}{1+\frac{2\mu\left(q_{1}-q_{0}^{d}\right)}{1+2\delta q}}\right)+\frac{1}{\tau}\int\mathcal{D}z_{0}\int\mathcal{D}u_{0}\log\int \mathcal{D}z\left[\frac{\cosh\left(\phi+2\mu\frac{\varrho(z_0, z, u_0, q_0^0, m, q_0^d)}{\varpi(\delta q, q_{1}, q_{0}^{d})/\sqrt{\tilde{q}}u_{0}}\right)}{e^{\mu\frac{\tilde{q}u_{0}^{2}+\varrho(z_0, z, u_0, q_0^0, m, q_0^d)^{2}}{\varpi(\delta q, q_{1}, q_{0}^{d})}}}\right]^{\tau},
        \label{eq:1RSB}
    \end{aligned}
\end{gather}
where

\begin{align*}
    \varpi(\delta q, q_{1}, q_{0}^{d})&=\left(1+2\delta q\right)+2\mu\left(q_{1}-q_{0}^{d}\right),\\
    \varrho(z_0, z, u_0, q_0^0, m, q_0^d) &= z_{0}\sqrt{q_0^0-m^{2}}+z\sqrt{q_0^d-q_{0}}+mu_{0}.
\end{align*}

Eq.\eqref{eq:1RSB} can now be used to investigate the RS solutions' global stability. They are globally stable if the RS saddle point is a stable fixed point of the 1RSB free entropy with unitary Parisi parameter. This finally indicates that RS solutions are not globally stable \cite{barbier2024phasediagramcompressedsensing}. 

\subsubsection{Local stability of RS solutions}
\label{sec:RSStability}

In the previous section, we stated that RS solutions are not global optimizers of free entropy $\Phi$. However, this does not prevent them from being locally stable. In this section, we address their local stability.

Local stability is typically assessed by evaluating the eigenvalues of the Hessian matrix of the free energy, whose positivity guarantees the local stability of the solutions \cite{Stability_Urbani}. As reported in \cite{ASP}, we can define two pairs of constants: two associated with the entropic contribution of $\Phi$ to the free entropy (labeled by S) and two corresponding to the energetic contribution (labeled by E).

\begin{gather}
    \begin{aligned}
        \lambda_{I}^{S}&=\int\mathcal{D}x_{0}\int \mathcal{D}\left(\frac{z_{0}-\hat{m}x_{0}}{\sqrt{\hat{q}_{0}}}\right) \left\{ \left(\partial_{z_{0}}^{2}-\left(\partial_{z_{0}}\right)^{2}\right)\log\int\mathcal{D}\left(\frac{z_{1}}{\sqrt{\hat{q}_{1}-\hat{q}_{0}}}\right)e^{\varphi^{in}\left(z_{0}+z_{1},\delta\hat{q}\right)}\right\} ^{2},\\\lambda_{II}^{S}&=\int\mathcal{D}x_{0}\int\mathcal{D}\left(\frac{z_{0}-\hat{m}x_{0}}{\sqrt{\hat{q}_{0}}}\right)\frac{\int\mathcal{D}\left(\frac{z_{1}}{\sqrt{\hat{q}_{1}-\hat{q}_{0}}}\right)e^{\varphi^{in}\left(z_{0}+z_{1},\delta\hat{q}\right)}\left\{ \left(\partial_{z_{1}}^{2}-\left(\partial_{z_{1}}\right)^{2}\right)\varphi^{in}\left(z_{0}+z_{1},\delta\hat{q}\right)\right\} ^{2}}{\int\mathcal{D}\left(\frac{z_{1}}{\sqrt{\hat{q}_{1}-\hat{q}_{0}}}\right)e^{\varphi^{in}\left(z_{0}+z_{1},\delta\hat{q}\right)}},\\\lambda_{I}^{E}&=\int\mathcal{D}u_{0}\int\frac{dz_{0}}{\sqrt{2\pi\left(q_{0}\right)}}e^{-\frac{z_{0}^{2}}{2\left(q_{0}-m^{2}\right)}}\left\{ \left(\partial_{z_{0}}^{2}-\left(\partial_{z_{0}}\right)^{2}\right)\left(\frac{1}{s}\log\left[\int\mathcal{D}\left(\frac{z_{1}}{\sqrt{{q}_{1}-{q}_{0}}}\right)e^{s\varphi^{out}\left(z_{0}+z_{1},\delta q,u_{0}\right)}\right]\right)\right\} ^{2},\\\lambda_{II}^{E}&=\int\mathcal{D}u_{0}\int\frac{dz_{0}}{\sqrt{2\pi\left(q_{0}\right)}}e^{-\frac{z_{0}^{2}}{2\left(q_{0}-m^{2}\right)}}\frac{\int\mathcal{D}\left(\frac{z_{1}}{\sqrt{{q}_{1}-{q}_{0}}}\right)e^{s\varphi^{out}\left(z_{0}+z_{1},\delta q,u_{0}\right)}\left\{ \left(\partial_{z_{1}}^{2}-\left(\partial_{z_{1}}\right)^{2}\right)\varphi^{out}\left(z_{0}+z_{1},\delta q,u_{0}\right)\right\} ^{2}}{\int\mathcal{D}\left(\frac{z_{1}}{\sqrt{{q}_{1}-{q}_{0}}}\right)e^{s\varphi^{out}\left(z_{0}+z_{1},\delta q,u_{0}\right)}},
    \end{aligned}
\end{gather}

where

\begin{gather*}
    \begin{aligned}
        \mathcal{D}\left(\frac{z-z_{0}}{\sqrt{\sigma}}\right) &= \int_{-\infty}^{\infty} \frac{e^{\frac{\left(\frac{z-z_{0}}{\sqrt{\sigma}}\right)^2}{2}}}{\sqrt{2\pi\sigma}}dz.\\        \varphi^{in}\left(z_{0}+z_{1},\delta\hat{q}\right)&=\frac{\left(z_{1}+z_{0}\right)^{2}}{2\delta\hat{q}},\\        \varphi^{out}\left(z_{0}+z_{1},\delta q,u_{0}\right)&=\log\left(\sum_{s=\pm1}e^{-S\phi}e^{-\mu\left[\frac{\left(\sqrt{\tilde{q}}u_{0}+S\left(z_{0}+mu_{0}+z_{1}\right)\right)^{2}}{1+2\delta q}\right]}\right).
    \end{aligned}
\end{gather*}

After some algebraic manipulations, it is possible to simplify the expression for entropic and energetic constants. 
\begin{align}
    \lambda_{I}^{S}=\frac{\delta\hat{q}^{2}+3\hat{m}^{4}+2\hat{q}_{0}^{2}+\hat{q}_{1}^{2}-2\delta\hat{q}\left(\hat{m}^{2}+\hat{q}_{1}\right)^{2}+2\left(\hat{q}_{1}+2\hat{q}_{0}\right)\hat{m}^{2}}{\left(\delta\hat{q}-\hat{q}_{1}+\hat{q}_{0}\right)^{4}},\\
    \lambda_{II}^{S}=\frac{6}{\delta\hat{q}^{2}}-8\frac{\delta\hat{q}+\hat{m}^{2}+2q_{0}-q_{1}}{\delta\hat{q}\left(\delta\hat{q}+q_{0}-q_{1}\right)^{2}}+3\frac{\left(\delta\hat{q}+\hat{m}^{2}+2q_{0}-q_{1}\right)^{2}}{\left(\delta\hat{q}+q_{0}-q_{1}\right)^{4}},
\end{align}

\begin{align}
\lambda_{I}^{E}&=\int\mathcal{D}u_{0}\int\mathcal{D}z_0\times\left\{ \left(\partial_{z_{0}}^{2}-\left(\partial_{z_{0}}\right)^{2}\right)\left(\log\left[\int\mathcal{D}z_1 \frac{\cosh\left(\phi+\mu\frac{2\varrho(z_0, z_1, u_0, q_0, m, q_1)\sqrt{\tilde{q}}u_{0}}{1+2\delta q}\right)}{e^{\mu\frac{\varrho(z_0, z_1, u_0, q_0, m, q_1)^{2}}{1+2\delta q}}}\right]\right)\right\} ^{2},\\
\lambda_{II}^{E}&=\int\mathcal{D}u_{0}\int\mathcal{D}z_{0}\frac{\int\mathcal{D}z_{1}e^{s\varphi^{out}\left(\vartheta(z_0, z_1, q_0, m, q_1),\delta q,u_{0}\right)}Arg\left(\vartheta(z_0, z_1, q_0, m, q_1),\delta q,u_{0}\right)^{2}}{\sqrt{\frac{q_{0}}{q_{0}-m^{2}}}\int\mathcal{D}z_{1}e^{s\varphi^{out}\left(\vartheta(z_0, z_1, q_0, m, q_1),\delta q,u_{0}\right)}},
\end{align}
where
\begin{align*}
\vartheta(z_0, z_1, q_0, m, q_1) &=z_{0}\sqrt{q_{0}-m^{2}}+z_{1}\sqrt{q_{1}-q_{0}},  \\
\varrho(z_0, z_1, u_0, q_0, m, q_1) &= \vartheta(z_0, z_1, q_0, m, q_1)+mu_{0},\\
Arg&=\left(\partial_{z_{1}}^{2}-\left(\partial_{z_{1}}\right)^{2}\right)\varphi^{out}\left(z_{0}+z_{1},\delta q,u_{0}\right).
\end{align*}

These constants are used for constructing a function of the eigenvalues of the hessian of $\Phi$: $\lambda_{I}^{E}\cdot\lambda_{I}^{S}$ and $\lambda_{II}^{E}\cdot\lambda_{II}^{S}$. If they are both smaller than the unity, we have local stability of the solution. Stability results are presented in Sec.\ref{subsec:stability}

\subsubsection{Local Stability of the RS Solution results}
\label{subsec:stability}

As previously announced in Sec.\ref{subsec:stability}, the RS solutions are not Globally stable. However, as it is possible to detect from the following figure, there is a regime in which they are locally stable.

\begin{figure}[H]
    \centering
    \includegraphics[scale=0.3]{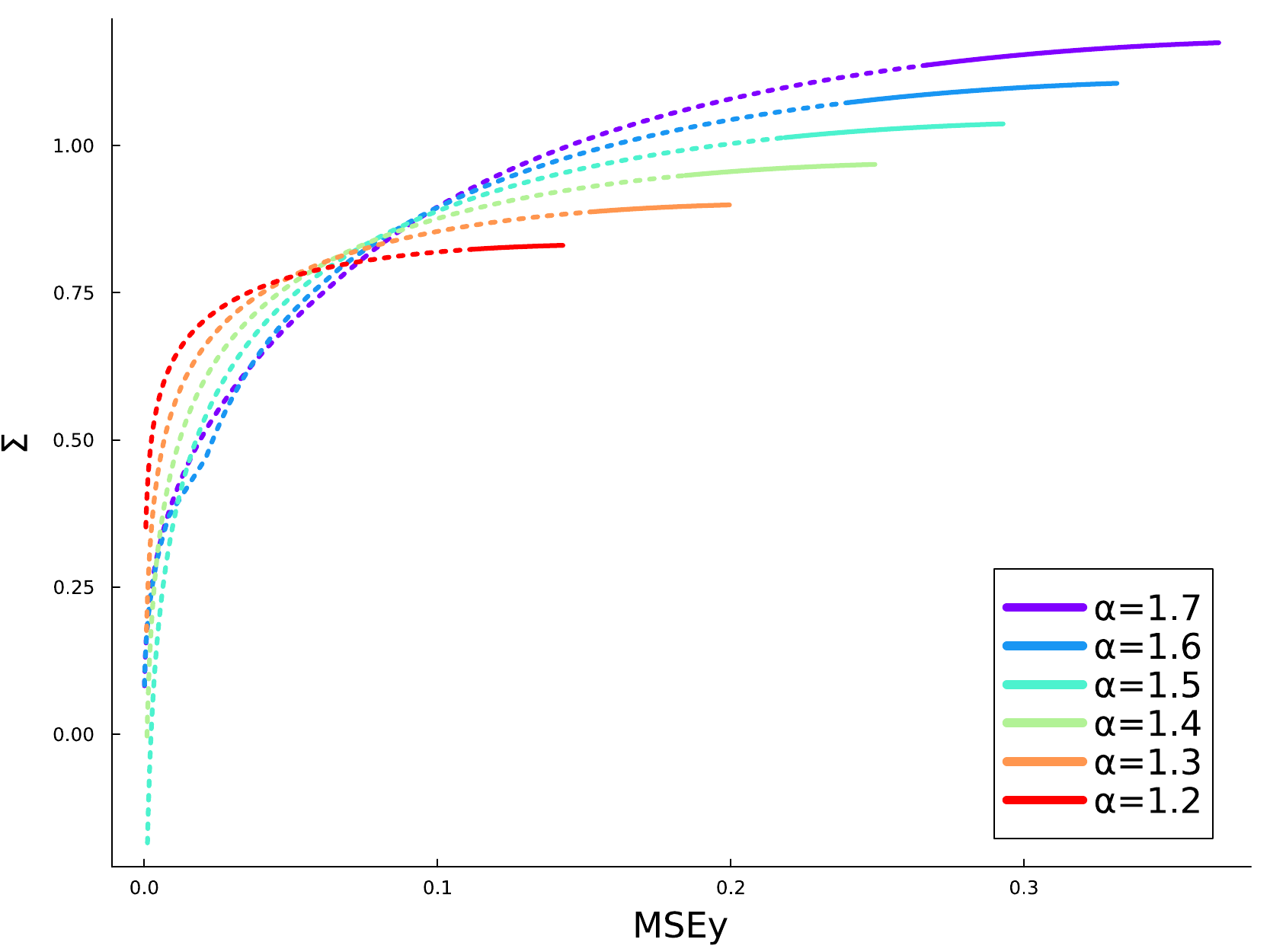}
    \caption{\textit{Left panel} Mean square error over labels $MSE_y$ vs $\Sigma$ at fixed $\lambda=0.0$ and values of $\alpha \in \{1.2,1.3.,1.4,1.5,1.6,1.7\}$. The locally unstable branch is a dashed line, the locally stable one is a solid line. }
    \label{Fig:Stability}
\end{figure}

It is evident from Fig.\ref{Fig:Stability} that the high complexity region of the branch remains locally stable, for the two different values of $\lambda$. This observation is consistent with the findings reported in \cite{Cui2020LargeDF}. The regularization effect is to locally stabilize regions characterized by smaller mean squared errors on the labels.

\section{Numerics}
\label{app:Numerics}
\subsection{Problem generator}
\label{app:ProblemGenerator}

In order to deal with numerical investigation a Phase Selection instances generator has been implemented, it follows in Alg.\ref{alg:Problem}.

\begin{algorithm}[H]
    \caption{Phase Selection instances generator}
    \label{alg:Problem}
    \scriptsize
    \begin{algorithmic}[H]
        \State \textbf{Input}: Structure: $N, \alpha, \text{act}, \Delta_0, P_0(\cdot), V_{\mathbf{x_0}}, \rho$\\
    
    \State \textbf{Initialization:} $M = \alpha N, M_{tst} = max\{200, M/4\}$\\

    \State \textbf{if} $P_0(\cdot) = \mathcal{N}(\cdot, \mathbf{0}, V_{\mathbf{x_0}})$
        \begin{align*}
            \mathbf{r,u} &\sim \mathcal{N}(\mathbf{0},I_N)\\
            x_{0,i} &= \sqrt{N V_{\mathbf{x_0}}} \frac{r_i}{\lVert \mathbf{r}\rVert_2} \cdot I\left(u_i < \rho\right)
        \end{align*}
    \\
    \hspace{1em}
    
    \State \textbf{Instance construction:} Contruction of $A, \mathbf{x_0}$ and $\mathbf{y}$

    \begin{center}
    \begin{minipage}{0.4\textwidth}
    \begin{align*}
        A &\sim \frac{\mathcal{N}_{N\times M}(\mathbf{0}, I_{N\times M})}{\sqrt{N}}\\
        \mathbf{v} &\sim \mathcal{N}_{M}(\mathbf{0}, I_{M})\\
        \mathbf{y} &= act(A \cdot \mathbf{x_0}) + \sqrt{\Delta_0} \mathbf{1}\cdot \mathbf{v}
    \end{align*}
    \end{minipage}
    \begin{minipage}{0.4\textwidth}
    \begin{align*}
            A_{tst} &\sim \frac{\mathcal{N}_{N\times M_{tst}}(\mathbf{0}, I_{N\times M})}{\sqrt{N}}\\
            \mathbf{v}_{tst} &\sim \mathcal{N}_{M_{tst}}(\mathbf{0}, I_{M_{tst}})\\
            \mathbf{y_{tst}} &= act(A_{tst} \cdot \mathbf{x_0}) + \sqrt{\Delta_0} \mathbf{1}\cdot \mathbf{v_{tst}}  
    \end{align*}
    \end{minipage}
    \end{center}
    \\
    \State \textbf{Return:} $A, \mathbf{y}, A_{tst}, \mathbf{y}_{tst}$
    \end{algorithmic}
\end{algorithm}

\end{document}